\documentclass[preprint,journal]{lib/ISMAR-TVCG-Template/vgtc}            %

\onlineid{2335}

\vgtccategory{Research}

\title{Looking Around by Looking Around: Omnidirectional Gaze-based VR Viewport Control}

\author{%
  \authororcid{Hock Siang Lee}{0000-0001-6263-6811},
  \authororcid{Jinghui Hu}{0000-0002-3965-2474},
  \authororcid{Florian Weidner}{0000-0001-8677-3503},
  \authororcid{Haopeng Wang}{0000-0003-2002-5216},
  and \authororcid{Hans Gellersen}{0000-0003-2233-2121}
}

\authorfooter{
    \item
  	Hock Siang Lee is with Lancaster University.
  	E-mail: h.s.lee3@lancaster.ac.uk
    \item
  	Jinghui Hu is with Lancaster University.
  	E-mail: j.hu23@lancaster.ac.uk
    \item
  	Florian Weidner is with University of Glasgow.
  	E-mail: florian.weidner@glasgow.ac.uk
    \item
  	Haopeng Wang is with Lancaster University.
  	E-mail: h.wang73@lancaster.ac.uk
    \item
  	Hans Gellersen is with Lancaster University and Aarhus University.
  	E-mail: h.gellersen@lancaster.ac.uk
}

\abstract{%
Traditional VR viewport control primarily relies on head and torso movement, which can be effortful and limiting in both constrained and extended-use settings. 
We introduce Looking Around by Looking Around (LALA), a gaze-based VR pitch-and-yaw viewport control technique designed for natural and effortless omnidirectional exploration via eye movements, without requiring or obstructing movement of the head, hand, or body, offering a low-effort and highly accessible interaction method.
Because gaze is primarily used for perception and exhibits oculomotor and perceptual asymmetries, using it directly for control is difficult.
To address this, we designed an asymmetric omnidirectional control profile for the eye, then built on it to exploit tendencies for eyes to stay within comfortable regions for viewport control.
We evaluated LALA in a user study (N=18) featuring two contrasting tasks: alignment towards known directions and open-ended visual search towards unknown directions.
LALA was strongly preferred over the traditional baseline, achieving competitive performance while enabling fully hands-free interaction with minimal physical movement.
}


\graphicspath{{lib/ISMAR-TVCG-Template}{lib/ISMAR-TVCG-Template/figs}{figs/}{figures/}{pictures/}{images/}{./}} %

\usepackage{times}                     %

\usepackage{subfig} %
\usepackage{enumitem} %
\usepackage{soul} %
\usepackage{multirow}
\usepackage{graphicx}
\usepackage{mathptmx}                  %
\usepackage{amsmath}
\usepackage{algorithm}
\usepackage{algpseudocode}
\usepackage{comment}
\usepackage[english]{babel}
\usepackage[autostyle, english = american]{csquotes}
\MakeOuterQuote{"}

\usepackage[size=tiny, colorinlistoftodos]{todonotes} %
\definecolor{ruby}{HTML}{9b111e}

\newcommand{\artcyay}[5]{$\hat{\Delta}$ = #1, SE = #2, \textit{t}(#3) = #4, \textit{p} $<$ #5}

\begin{document}

\maketitle

\section{Introduction}

Virtual reality (VR) enables users to explore immersive environments by naturally looking around. 
In most VR systems, viewport orientation is directly controlled through one-to-one mapping between head pose and the virtual camera. 
While this mapping preserves a strong sense of spatial correspondence between physical and virtual motion, it shifts the burden of comfortable interaction onto the users.
In conventional VR, users are often expected to be able to freely rotate their head and torso to explore the surrounding environment, or rely on external accommodations, such as swivel chairs or unrestricted space, to remain comfortable.
This assumption breaks down in many real-world situations where large movements are impractical, such as when users are seated, lying down, traveling, or in public settings where extensive movement may be constrained or socially awkward.
As a result, not only is VR often fatiguing and uncomfortable, but its broader adoption is constrained by where, for how long, and by whom it can be comfortably used.

To reduce the need for external accommodations and large physical movements, researchers have explored techniques that decouple viewport rotation from head motion.
Controller-based techniques allow users to rotate the view using handheld devices, but they require manual input and occupy the hands, making them less suitable for hands-free interaction~\cite{Skalski2011FunNaturalControllerMapping}.
Head amplification techniques instead increase the virtual rotation produced by head movement, reducing the amount of physical motion required.
However, amplification introduces a trade-off between comfort and controllability: Low gain provides only a limited reduction in physical effort. High gains reduce physical effort but increase sensitivity, causing large virtual rotations even with small or unintended head movements such as overshooting, disorientation, and simulator sickness~\cite{Wang2023RotationGainsPerceptual}. 
This can make the viewing experience unstable and reduce the sense of spatial control.

In this work, we explore an alternative perspective motivated by the idea of Looking Around by Looking Around (LALA), which extends the viewport through natural eye movements.
During visual exploration, the eyes typically move first to inspect peripheral regions before the head follows to bring those regions into central view. 
This eye-head coordination suggests that gaze can serve as a natural signal for extending the viewport.
However, because gaze is primarily used for perception rather than control, designing a purely gaze-driven viewport control technique requires balancing responsiveness with perceptual stability.

Hence, the design of LALA was carefully guided by three principles aimed at ensuring user experience and comfort: (1) preserving stable viewing during natural gaze behavior, (2) enabling smooth and continuous viewport control for exploration, and (3) supporting rapid reorientation during large gaze shifts.
To achieve this, LALA converts gaze signals into a control profile derived from the physiological and behavioral properties of human eye movements, and integrates this profile into a viewport control mechanism that accounts for both \textit{oculomotor} and \textit{perceptual} asymmetries.

We first construct an asymmetric omnidirectional gaze control profile that maps eye-in-head positions to control activation values (\autoref{subsec:Techniques:eyeModel}).
We then translate these activation signals into viewport control formulations for yaw (\autoref{subsec:LALAyaw}) and pitch (\autoref{subsec:LALApitch}) to account for perceptual asymmetries.
Finally, as rotations are now decoupled from the HMD, we remap rotations and head movements to maintain natural viewing (\autoref{subsec:remappingHeadMovements}).

Together, these mechanisms of LALA produce an asymmetric omnidirectional control profile that enables natural, hands-free exploration while maintaining stable and comfortable viewing, allowing the eyes to steer the viewport toward any direction without constraining head movement.
To our knowledge, LALA is the first to enable intuitive gaze-only pitch–yaw viewport control, achieved by accounting for oculomotor and perceptual asymmetries.

We evaluate LALA through controlled user studies comparing it with head amplification and baseline 1:1 mapping. 
Our results show that LALA achieved performance competitive to existing techniques while enabling fully hands-free interaction with minimal physical movement.
Participants also strongly preferred LALA over our baselines, reporting favorable subjective experiences when using LALA to explore the virtual environment.

This work contributes: 
\begin{itemize}[noitemsep]
    \item A gaze-based viewport control technique that enables stable omnidirectional VR exploration through natural eye movements alone, without requiring or obstructing movement of the head or torso.
    \item A control profile design for the eyes that accounts for both oculomotor and perceptual asymmetries and characteristics.
    \item An empirical evaluation demonstrating that LALA can support practical exploration of VR environments with strong user preference and qualitative results better or competitive to existing head-based techniques.
\end{itemize}

\section{Viewport Control Techniques}\label{sec:rw}
VR is often experienced without visibility of the physical environment and within limited real-world space, making intuitive and space-efficient travel techniques essential.
These techniques include viewport control (controlling the virtual FOV) and locomotion (translating in virtual space), though research has focused more heavily on locomotion~\cite{Di2021LocomotionVault} as it is commonly assumed that users have enough space to control their view through rotation of head and torso.
However, the range of head movement can be limited by posture~\cite{Mcgill2020SeatedWorkspaces, Sidenmark2019EyeHeadTorsoDuringGazeShift,Gemert2023Bed}, situation~\cite{Eghbali2019acceptability,Bajoranaite21-VRpublictransport} or ability~\cite{Mott2020VRaccessibility,Hansen2019VrWheelchair}.

Alternative viewport control techniques are often designed to reduce user effort while improving navigation and accessibility~\cite{mine1995virtual}, with a focus on modifying the horizontal yaw rotation.
Within academia, head amplification techniques are a popular hands-free alternative, involving amplifying the movements of the head for users to reach further with head movement alone than physically possible~\cite{Westhoven2016HeadAmplification, Sargunam2017HeadRedirectionGuided&Amplified, Jay2003HeadAmpliciation}.
Amplifying by a constant gain factor retains an absolute mapping between head pose and the viewport, supporting a user's sense of spatial orientation~\cite{Ragan2017AmplifiedHead}, but dynamic gain has also been explored~\cite{Langbehn2019TurnYourHeadHalfRound,Lim2025CaliViewContinuousViewpoint,Zhang2021VelocityguidedHeadAmplification}.
A concern with head amplification is VR sickness due to the difference in visual and vestibular movement~\cite{Norouzi2018amplHead-VRsickness}.

VR viewport control has often been explored using external controllers and sensors too, utilizing various body parts as a means of control.
The most common techniques involves buttons, joysticks, or similar inputs on hand-held controllers ~\cite{Sargunam2018EvalVRRotationWithJoystick}.
Hands-free techniques that use body-leaning~\cite{Marchal2011HumanJoystick, Sato2015VibroSkate, Vezzani2023WalkingSeat}, shoulder-rotation~\cite{Guy2015LazyNav}, arm-movement~\cite{McCullough2015VRArmSwingLocomotion} and more can also be found, but these approaches are typically designed for locomotion tasks.

There are also techniques that expand the user’s effective FOV without interaction by using visual strategies such as scaling panoramic views to fit within a narrow FOV~\cite{Ardouin2012FlyViz}, layering multiple perspectives~\cite{Schjerlund2022OVRlap}, or introducing pictures-in-picture~\cite{Lin2017OutsideInVisualizingOutofSight}.

Most recently, with eye-tracking now increasingly available in modern VR HMDs, academia have also recently started exploring using gaze for viewport control. 
Most notable for our work is how Lee et al.'s Gaze Gain and Gaze Pursuit techniques~\cite{Lee2024SnapPursuitGainViewportControl} harnesses different eye movements, even though they are 1D yaw-only techniques prioritizing performance (as opposed to comfort) and cannot be used for combined pitch-and-yaw rotations.
Gaze Pursuit leverages smooth pursuit eye movements through a relative mapping, continuously rotating the viewport in the direction of the eye, with the rotation speed proportional to the eye-in-head angle.
In contrast, Gaze Gain leverages saccadic eye movements with an absolute mapping, rapidly rotating the viewport toward an amplified combination of head and eye yaw angle, with rotation speed governed by a transfer function~\cite{Lee2024SnapPursuitGainViewportControl}.
Both of these techniques utilized learnings from general gaze-based navigation, including scrolling and panning on desktops~\cite{Sharmin2013GazeScrolling} and public displays~\cite{zhang13-sideways,Zhang2015GazeHorizon}, and for 2D camera control in 3D video games~\cite{Badler2015GazeGameNavigation, Velloso2016EyePlay}. 
Though, works that have used gaze to manipulate the view in VR also exist, for example enabling gaze to orbit around an object~\cite{Pai2017GazeSphere}, to subtly move the viewport during saccades for redirected walking~\cite{Sun2018RedirectedWalkingDuringSaccades}, or to use peripheral gaze as a trigger for torso rotation~\cite{Lee2024RPGRotationTechnique}. 
Importantly, we highlight that extending gaze-based viewport control from yaw to omnidirectional pitch–yaw is not a straightforward dimensional extension, requiring a redesign of the control model itself (see \autoref{subsec:Techniques:eyeModel}).
Simple extensions would be difficult to use due to fundamental oculomotor (see \autoref{subsec:Techniques:eyeModel}) and perceptual asymmetries (see \autoref{subsec:LALApitch}) between horizontal and vertical gaze.

\section{LALA Design}\label{sec:techniques}
LALA is an omnidirectional gaze-based viewport control technique that allows users to stably explore VR environments by inducing additional viewport rotation toward the direction of gaze.
Although gaze naturally reflects visual attention and user intent, the human visual system is optimized for perception rather than control.
Consequently, effective gaze-based viewport control must account for the characteristics and limitations of eye movements and visual perception.

An important consideration in omnidirectional viewport control is the oculomotor and perceptual asymmetries between the horizontal and vertical axes.
\textit{Oculomotor asymmetries} refer to the direction-dependent characteristics of eye movements, such as different directions requiring differing amount of effort.
\textit{Perceptual asymmetries} refer to our familiarity perceiving relative to a gravity-aligned reference frame and that vertical movements interact more strongly with gravitational and vestibular cues.

LALA accommodates for oculomotor asymmetries a asymmetric omnidirectional gaze control profile that maps eye-in-head positions to control activation values (\autoref{subsec:Techniques:eyeModel}).
We then translate these activation signals into distinct viewport control formulations for yaw (\autoref{subsec:LALAyaw}) and pitch (\autoref{subsec:LALApitch}) to account for oculomotor asymmetries.
Finally, as rotations are now decoupled from the HMD, we remap rotations and head movements to maintain natural viewing (\autoref{subsec:remappingHeadMovements}).

Notably, complexities of the human visual and oculomotor systems heavily influence LALA's design.
While these complexities are well studied in vision science, they are less familiar in HCI and VR interaction research, so we detailed them together with the designs of LALA (as opposed to a related work section).
We highlight that explicit use of vision science as a basis for interaction design is rare in HCI, and can be considered part of the contribution of LALA.

\subsection{Asymmetric
Omnidirectional Gaze Control Profile}\label{subsec:Techniques:eyeModel}
LALA converts gaze data into an asymmetric omnidirectional control profile allowing easy interpretations of gaze signals, serving as the foundation for downstream viewport control.
The goal of this profile is to capture the key characteristics of human eye movements so that gaze signals can be interpreted in a way that supports stable viewing, smooth continuous control and rapid reorientation.
To achieve this, the control profile is derived from the behavioral and physiological properties of fixations, smooth pursuit, and saccadic eye movements.

To design such a profile, we first consider how gaze has been used in general interactions.
Large eye-in-head angles have been used as an indicator of the user's intent to interact with interface elements, as they often reflect effortful gaze shifts since natural gaze rarely enters such peripheral regions~\cite{Choi2022KuiperBelt, Sidenmark2019EyeHeadTorsoDuringGazeShift}.
These interactions often use a trigger-based binary decision.
However, smooth viewport control requires a continuous control signal to avoid jerky start–stop behavior caused by eye movements such as the optokinetic reflex.
This control signal is most naturally defined as an absolute mapping of eye-in-head angles, rather than relative velocity or gestures. 
Eye positions inherently encode spatial offsets, which users can interpret through proprioception and their understanding of the surrounding environment, making the relationship between eye angle and control signal more intuitive and easier to learn compared to relative mappings.

Having established that viewport control should be driven by a continuous mapping of eye-in-head angles, the next question is how various eye movements affect this mapping.
Fixations, which typically occur near the central eye-in-head region, involve aligning a stationary object or area of interest with the fovea for at least $\sim$200 ms to enable detailed visual processing~\cite{Land2012Looking&Acting}.
Hence, to maintain stable viewing within the central eye-in-head region, the viewport should remain largely stationary, corresponding to a control signal close to zero.

Outside of this central region, while gaze remains comfortably reachable by the eyes alone, scene rotations induced by viewport control would cause fixations to naturally transition into smooth pursuits.
Smooth pursuits are particularly important for viewport control, as they enable users to continuously track objects and maintain engagement with displayed content while the viewport is moving. 
However, smooth pursuit can reliably track motion only up to approximately $\sim$30\textdegree/s before catch-up saccades are triggered~\cite{Meyer1985UpperLimitSmoothPursuit}. 
While saccades are much faster, their ballistic nature makes their endpoints almost impossible to adjust once initiated~\cite{Kornylo2003CancelingPursuitAndSaccadicEyeMovements}.
Consequently, the ideal control signal for viewport rotational speeds must be constrained while within this region.
If the viewport moves too quickly, smooth pursuit will frequently trigger catch-up saccades, and saccades may miss their intended landing positions as visual targets shift during their execution. 
This results in viewport motion that appears stuttery and difficult to control~\cite{Lee2024SnapPursuitGainViewportControl}, potentially increasing cybersickness and disorientation~\cite{Stauffert2018EffectsOfLatencyJitter, Stauffert2020MotiontoPhotonOnCybersickness}.

Further out represents the peripheral eye-in-head region, which the eyes can reach but not comfortably linger in. 
This region is rarely accessed during natural gaze and is typically only reached during large-amplitude gaze shifts that are assisted by head movements~\cite{Sidenmark2019EyeHeadTorsoDuringGazeShift}. 
Such gaze shifts can even produce eye-in-head angles that exceed modern VR HMDs’ FOVs, for example, when tracking a fast moving object and our visual system predicts a future position outside the FOV.
Because these saccades are often guided top-down rather than toward a currently visible stimulus, the constraint limiting viewport speed to prevent saccades from missing its intended target is less critical.
Instead, the limiting factor for viewport speed is our ability to process newly appearing scene content and deciding when to stop or adjust gaze.
Consequently, the control signal in the peripheral region should increase substantially to enable rapid reorientation, but remain bounded to avoid overwhelming visual processing, which would degrading user control~\cite{Lee2024SnapPursuitGainViewportControl}.

Overall, the control signal can be conceptualized as having three regions where interactions are primarily from fixations, smooth pursuits, or saccades. 
It stays near zero in the central region to support stable fixations, rises gradually through the mid-range to accommodate smooth pursuits, and increases sharply in the peripheral region, where it is ultimately bounded by visual processing limits. 
To capture this behavior, we use a sigmoidal formulation, which ensures smooth transitions between regions while constraining the control signal at extreme eye-in-head angles.

\begin{figure}[tb]
    \centering
    \subfloat[]{\includegraphics[width=.48\linewidth]{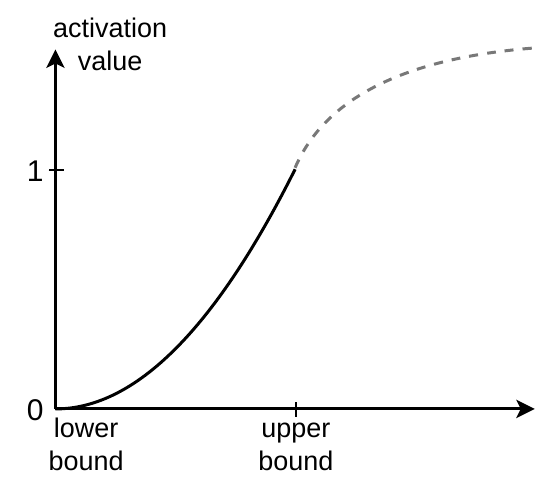}\label{fig:1dEyeModel}}\;
    \subfloat[]{\includegraphics[width=.48\linewidth]{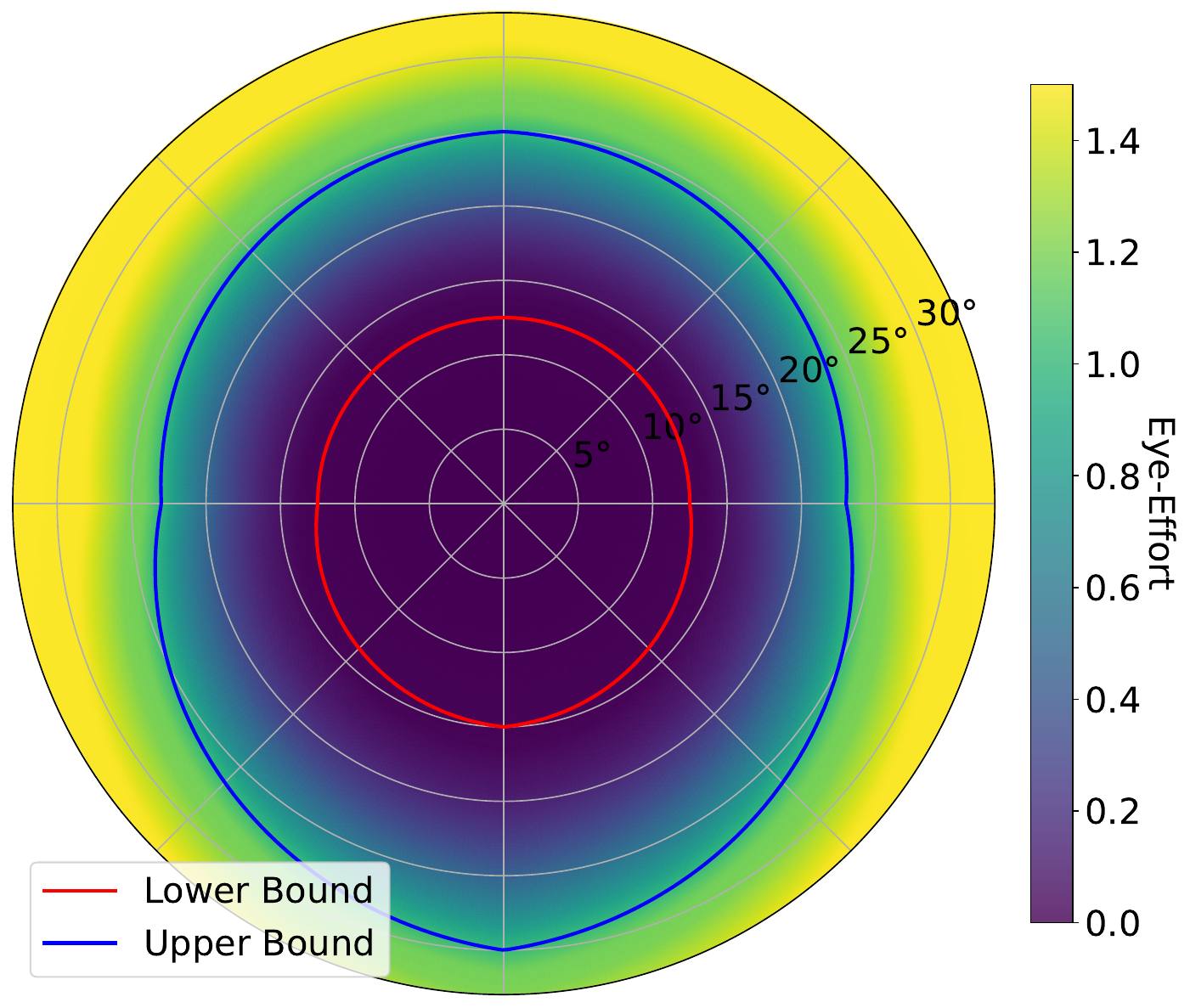}\label{fig:2dEyeModel}}\;
    \caption{(a) Activation curve between a lower and upper bound from a slice of (b) which is the full 2D mapping between eye-in-head directions and activation values that accounts for anatomical asymmetries of the eye across axes.}
    \label{fig:EyeModel}
\end{figure}

For the early and mid-range response of the sigmoid --- representing the central and still-comfortable regions of the eye --- we hypothesize that the muscular effort required to maintain peripheral eye positions correlates with the strength of the user's intent to interact with the viewport.
Accordingly, we base this portion of the curve on measurements of eye-muscle tension required to maintain \textit{horizontal} eye positions~\cite{Collins1975EyeMuscleTension}.
To ensure smooth behavior beyond this range ($>$1) --- representing the peripheral, sometimes still-reachable regions of the eye --- we extend the curve using Hermite interpolation as it accounts for the curve derivative, allowing the curve to more smoothly flatten.
We intentionally used values $>$1 so that the 0–1 range could remain well supported by vision science, whereas eye behavior at extreme angles is less understood.
Extending beyond this range also accommodates measurement noise, inter-individual variability, and eye-in-head angles that may exceed the HMD’s FOV.
Combined, this produces an overall sigmoid-like curve as seen in \autoref{fig:1dEyeModel}.
The lower asymptote at $y=0.01$ was motivated by pilot testing, which suggested that hard stops were less intuitive for some users, while the upper asymptote at $y=1.5$ was selected as a pragmatic extension.

However, oculomotor asymmetry prevents simple extrapolation of the control signal from horizontal-only to omnidirectional.
Our visual and oculomotor system does not treat the vertical pitch axis symmetrically with the horizontal yaw axis~\cite{Bonato2009Pitch&RollCybersickness, Kobel2021GravityMotionPerception, Grasse1992AnalysisNaturallyOccurring}.
For instance, smooth pursuits are inherently asymmetric, with performance generally superior for horizontal yaw compared to vertical pitch~\cite{Grasse1992AnalysisNaturallyOccurring}.
These oculomotor asymmetry affect eye-muscle tension, which in turn affects preferred eye-in-head amplitudes across gaze bearings.

To account for this, the control signal was scaled to fit between the lower and upper bounds of preferred eye-in-head amplitudes observed during gaze shifts towards different bearings~\cite{Sidenmark2019EyeHeadTorsoDuringGazeShift}.
We assume the overall shape of the sigmoid (i.e., the general relationship between eye-in-head angle and effort) is similar across bearings, and that direction-dependent differences are sufficiently accounted for by the directional bounds.
This results in the 2D pitch-and-yaw representation depicted in \autoref{fig:2dEyeModel}.
Together, these bounds create a 2D control profile that may be perceived as comprising three regions: a central area for stable viewing, an intermediate region supporting comfortable passive viewport control, and a peripheral zone enabling active viewport control.

The resulting activations values from the control profile is defined with respect to eye-in-head space, which can be decomposed into components along the relevant world axes using basic trigonometry.
This is particularly important as eye rotations do not align with world axes when head roll is present.
All in all, our control profile accounts for many of the complexities of the eye, which has the additional benefit of decoupling and simplifying downstream technique design from such complexities.
The representative values~\cite{Sidenmark2019EyeHeadTorsoDuringGazeShift, Collins1975EyeMuscleTension} are provided together with the source code in the Appendix.

\subsection{Perceptual Asymmetry: Differences in Pitch and Yaw Control}
The gaze control profile described above produces activation values in eye-in-head space, which must be translated into viewport rotations to enable gaze-driven navigation.
However, this translation cannot be applied uniformly across directions. 
Human perception treats horizontal (yaw) and vertical (pitch) rotations differently, largely due to gravity and vestibular cues. 
As a result, LALA adopts different control formulations for yaw and pitch: yaw is implemented using a relative control, while pitch is implemented using an absolute mapping aligned with gravity.

\subsubsection{Relative Yaw Control via Gaze}\label{subsec:LALAyaw}
For yaw, our goal is to enable easy 360\textdegree{} surround viewing.
Among the yaw-only gaze-based techniques reviewed in \autoref{sec:rw}, the Gaze Pursuit technique~\cite{Lee2024SnapPursuitGainViewportControl} achieved the strongest performance, so we draw inspiration from it.

In both LALA and Gaze Pursuit, the yaw of the viewport continuously rotates in the direction of the eye in the horizontal direction.
As an object of interest comes into view, the eye naturally fixates and smooth pursuits back towards the centre.
\begin{equation}\label{eq:LALAyawvelocity}
\begin{split}
\text{velocity}_{yaw} &= m \times \text{activationValue}_{horizontal} \\
    &\phantom{=} \times (\text{eyeInVRWorld}_{yaw} - \text{currentDirection}_{yaw})
\end{split}
\end{equation}

However, both techniques differ in how they determine the viewport angular velocity.
LALA's yaw velocity (see \autoref{eq:LALAyawvelocity}) is proportional to the activation value (see \autoref{subsec:Techniques:eyeModel}) and $\text{eyeInVRWorld}_{yaw}$ (the angle between the current viewport direction and the current gaze direction in VR).
$m$ is a tunable parameter controlling interaction sensitivity, analogous to cursor sensitivity in pointing interfaces, and was set to 4 based on pilot testing to achieve stable and usable behavior.

\subsubsection{Absolute Pitch Control via Gaze}\label{subsec:LALApitch}
For pitch, our goals are to enable easy and intuitive $\pm90$\textdegree{} pitch control.
However, unlike yaw, this is more complicated due to perceptual asymmetries.

During natural horizontal gaze shifts (yaw), gravity remains aligned with the head’s orientation, meaning gravity-sensitive receptors (e.g., the otoliths) are minimally affected. 
In contrast, pitch is more strongly affected by the continuous feedback from these gravity-sensitive receptors~\cite{Vidal2006YawAndPitchPerceptionCues, Kobel2021GravityMotionPerception}, as opposed to the current view direction.
This causes perceptual asymmetry between yaw and pitch, whereby yaw is perceived in a relative frame of reference, while pitch is perceived in an absolute frame of reference relative to gravity.

Hence, a naive relative mapping for pitch control is unintuitive. 
It makes all vertical motion equally effortless, allowing users to rotate past natural head movement limits --- looking fully up and down --- without resistance, easily rotating into being "upside-down". 
Yet, users are unlikely to want or benefit from such orientations, as humans have limited spatial familiarity and reduced sensory resolution when presented with upside-down sceneries~\cite{Kobel2021GravityMotionPerception, Cleary2014InversionEffects}. 
Therefore, viewport pitch control is ideally implemented using an absolute mapping, with reference to human-perceivable cues such as gravitational or proprioceptive signals.

\begin{equation}\label{eq:LALAtargetpitch}
\text{goalDirection}_{pitch} = g \times \text{eyeInWorld}_{pitch}
\end{equation}

In LALA, we employ an absolute mapping between viewport pitch and eye-in-world angle (\autoref{eq:LALAtargetpitch}) with an amplification factor $g = 2$ to obtain a "goalDirection", representing where the viewport should pitch to.
An amplification of 2 allows the technique to cover the entire [-90\textdegree{}, 90\textdegree{}] pitch range using only eye movements. 
Combined with the relative yaw mapping, it means upright viewing towards any direction is possible using only eye movements.

\begin{equation}\label{eq:LALApitchvelocity}
\begin{split}
\text{velocity}_{pitch} &= m \times \text{activationValue}_{vertical} \\
&\phantom{=} \times (\text{goalDirection}_{pitch} - \text{currentDirection}_{pitch})
\end{split}
\end{equation}

The viewport pitch velocity towards this "goalDirection" is governed by \autoref{eq:LALApitchvelocity}, where $m$ is a tuning constant controlling interaction sensitivity, set to 1.8 based on pilot testing to achieve stable and usable behavior.
This means LALA's pitch can be interpreted as a "delayed" absolute mapping, where its pitch velocity is proportional to both the activation value (see \autoref{subsec:Techniques:eyeModel}) and the angular difference between the current viewport pitch and the amplified gaze pitch.
A "delayed" approach is necessary here to prevent disorientation from saccades missing their intended landing positions if visual targets shift too much during their execution.

This design improves intuitiveness (by aligning viewport movement to gravity, allowing a "sense" of up/down during movement) while preventing users from unintentionally disorienting themselves or flipping the scene upside down (by using an absolute mapping for pitch tuned towards covering the $\pm$90\textdegree{} pitch range).
We also added a 1\textdegree{}/s rotation towards 0 pitch if the viewport pitch is within $\pm10$\textdegree{} to help realignment back to the horizon and sense of orientation.

Notably, the yaw and pitch velocity equations in LALA share the same structure, differing only in the reference direction used to compute the angular difference relative to the current viewport direction. 
This highlights the advantage of \autoref{subsec:Techniques:eyeModel}, which abstracts eye movement complexities from technique design.
This allows the remainder of LALA to remain agnostic to the underlying behaviour of the eyes, making a complete technique redesign between axis unnecessary.
Furthermore, both velocity formulas intentionally use homogeneous equations containing no constant terms, ensuring that small to moderate eye-in-head angles do not induce viewport rotations. 
In other words, while the eye remains within a comfortable range, the viewport behaves perceptibly the same as the default 1:1 HMD mapping, ensuring that LALA remains effectively invisible until an “uncomfortable” gaze shift occurs.

\subsection{Remapping Rotations and Head Movement}\label{subsec:remappingHeadMovements}

In LALA, viewport orientation is partially decoupled from the HMD because gaze can rotate the view independently of head motion. 
This decoupling introduces ambiguities in how head rotations should affect the final viewport rotation. 
Under standard quaternion composition, yaw and pitch interact depending on the current orientation of the viewport.
As a result, head rotations could produce different perceived motions depending on where the user is looking, leading to inconsistent yaw and pitch behaviour. To avoid this issue, we remap head and viewport rotations to preserve perceptually consistent control.

To address this, LALA (see \autoref{eq:LALApitchvelocity} and \autoref{eq:LALAyawvelocity}) uses an Euler-angle–like mapping which preserves the perceptual separation between yaw and pitch, which would otherwise break under quaternion composition.
The mapping intentionally re-introduced gimbal lock when looking directly up or down to reflect natural limits of human head rotation and visual search behavior under gravity. 
All head and eye yaw rotations are remapped to only alter the viewport (global) yaw regardless of current viewport pitch; and all head and eye pitch rotations are remapped to only alter the viewport (global) pitch regardless of the current viewport yaw.
This behavior is consistent with known constraints on biological eye–head coordination such as Listing’s law~\cite{Hestenes1994InvariantBodyKinematics, Wong2004ListingsLawClinical}.

This remapping better matches natural head movement and how rotations are interpreted in egocentric reference frames. 
For example, when looking upward while spinning in an office chair, perceived pitch (looking up/down) and yaw (turning left/right) remain consistent relative to the body and gravity-defined vertical. 
In contrast, aircraft-style rotation models tilt the entire reference frame, causing subsequent yaw rotations to behave differently. By aligning head and viewport movements with the body and gravity-aligned vertical, the interaction feels more intuitive~\cite{Alsmith2017HeadTorsoMisalignmentSpatialReferenceFrames}.

Likewise, head displacements are remapped to the current viewport orientation, so moving the HMD in the direction it faces moves the viewport in the direction it faces, regardless of their relative orientations. 
This allows users to, for example, gain elevation in VR by pitching the viewport upward and walking forward. 
Although our study did not elicit large head displacements, this remapping remained necessary because natural head and torso rotations produce small but non-negligible HMD and eye displacements.

In summary, LALA combines four modular components ---  asymmetric omnidirectional gaze control profile, a yaw control, a pitch control, and head-movement remapping --- to enable natural, omnidirectional viewport control with minimal physical effort. 
Together, these modules allow the technique to remain mostly unobtrusive during comfortable eye movements, while supporting intuitive navigation when larger gaze shifts occur. 

\section{Evaluation}
To evaluate our viewport control techniques across diverse conditions, we conducted a user study consisting of two tasks and compared them to existing baselines.
We selected these tasks as they represent contrasting user behaviors with distinct gaze patterns and cognitive demands: efficient target acquisition where the direction is known and predictable, versus open-ended visual search for targets at unknown and unpredictable directions.

The first is a modified Fitts’ law task (see \autoref{subsec:evaluation:task1design}) where participants looked for a target at a predictable and known direction, allowing them to move efficiently towards the current target while planning their movements accordingly for future targets.
The second is a visual search task (see \autoref{subsec:evaluation:task2design}) where participants freely searched their surroundings for a target at an unknown direction, necessitating them to scan their environment without prior knowledge of target locations.

In both tasks, targets are selected via eye-gaze combined with a manual button press, allowing participants to have a fast, immediate, explicit way to confirm when the target is perceived to have aligned with their gaze.
The design choice of using button selection as opposed to gaze-dwell selection, which is common in gaze-based interactions, is due to our focus on evaluating viewport alignment performance rather than selection mechanisms.
By requiring an intentional button press, we avoid the Midas Touch problem in which unintended gaze fixations trigger false selections, while also eliminating effects caused by different gaze dwell durations.

The targets appear 10m away from the participant and are 15 visual degrees in size (the approximate size of macular vision), making it easy to select even in the presence of eye-tracking inaccuracies. 
Combined with the button-press selection method, this design emphasizes viewport alignment --- bringing targets into view --- rather than precise selection accuracy.

\begin{figure}[tb]
    \centering
    \subfloat{\includegraphics[width=.4180\linewidth]{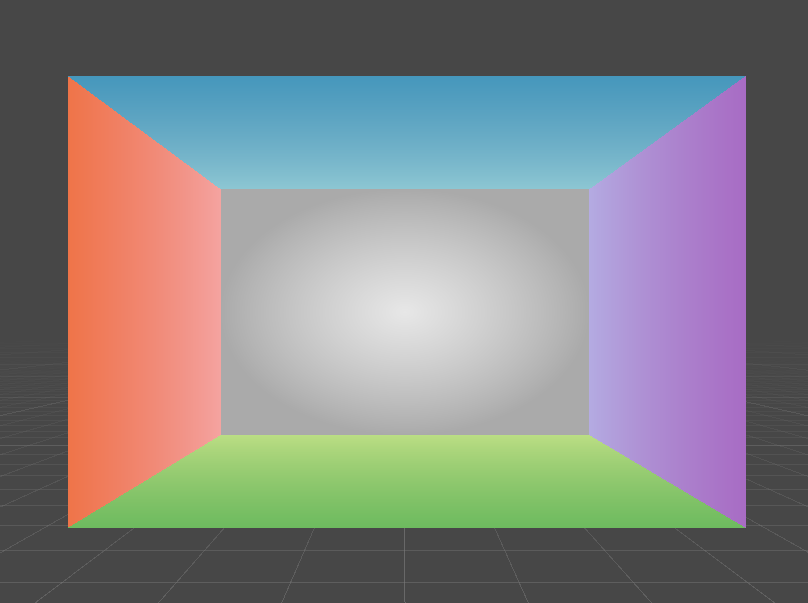}}\;
    \subfloat{\includegraphics[width=.3116\linewidth]{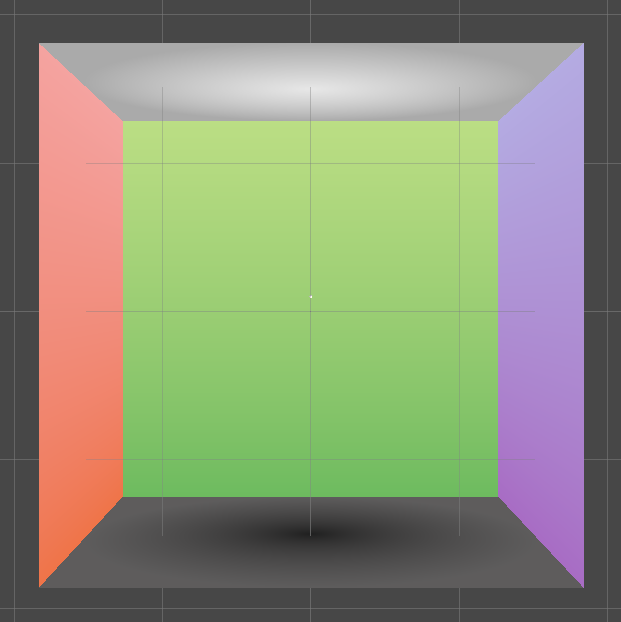}}\;
    \caption{Cube environment for both our tasks viewed from the front (left image) and top (right image). Image proportions modified for better sense of perspective and orientation.}
    \label{fig:FittsLawRoom}
\end{figure}

To give users a clear sense of orientation during both tasks -- and to avoid the perception that targets were moving rather than the viewport -- it was necessary to include sufficient visual cues in the environment. 
However, visual cues and distractions could introduce noise into the data. 
To balance these needs, both tasks were conducted in a large virtual cube (30×30×30m) with walls in distinct colors, each featuring a subtle gradient that becomes lighter in the forward-facing direction, as shown in \autoref{fig:FittsLawRoom}.

\subsection{Alignment Task: Looking towards known directions}\label{subsec:evaluation:task1design}
\begin{figure}[tb]
    \centering
    \subfloat[]{\includegraphics[width=.3\linewidth]{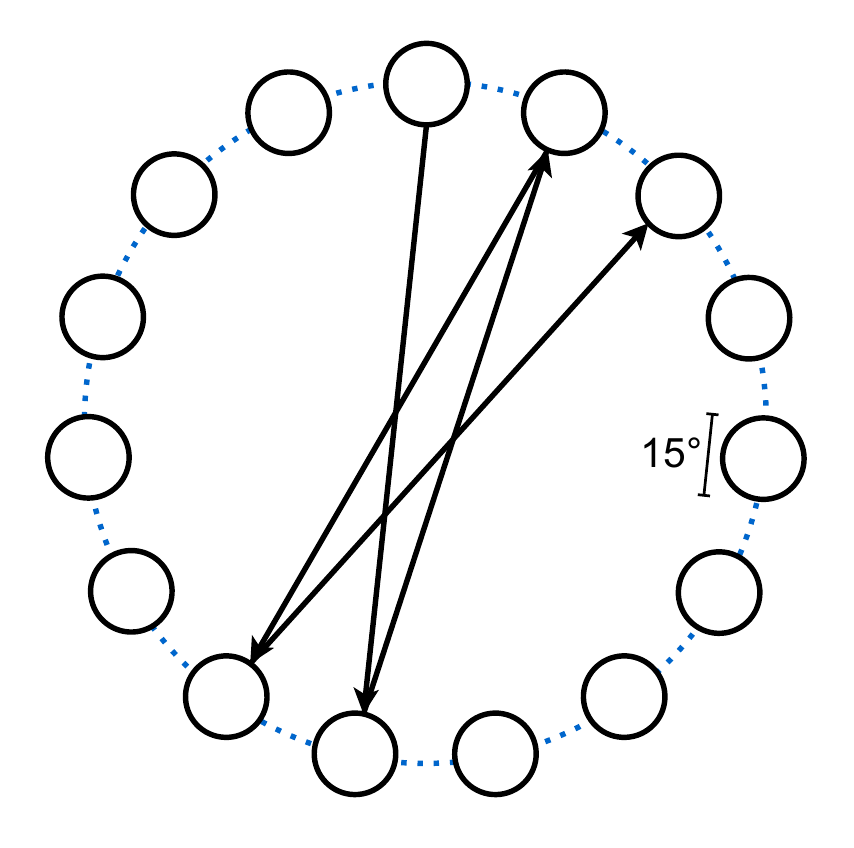}}\;
    \subfloat[]{\includegraphics[width=.3\linewidth]{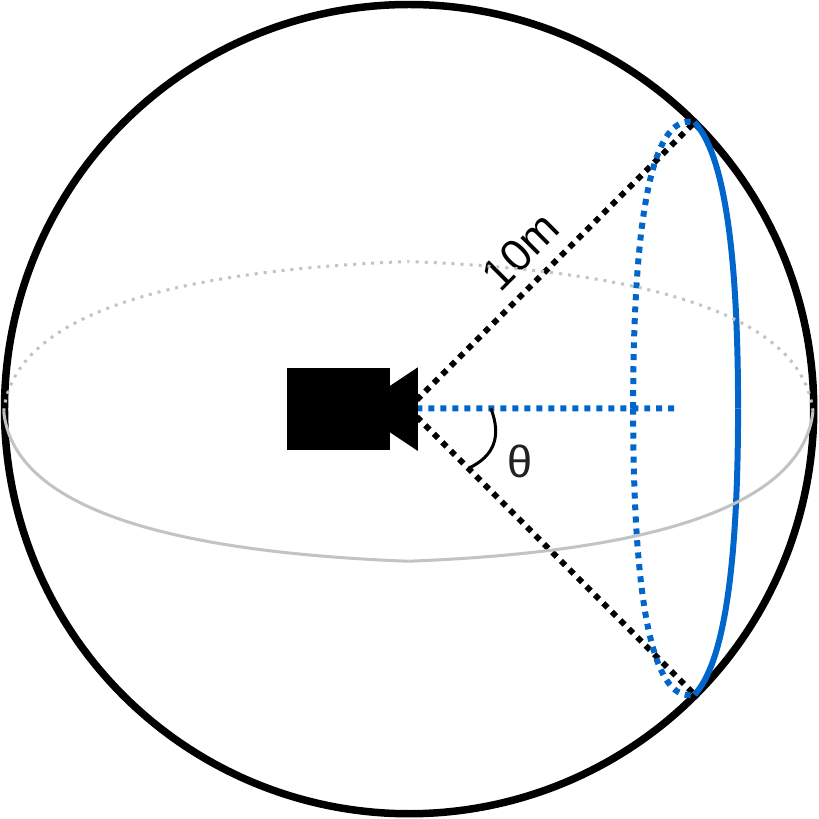}}\;
    \subfloat[]{\includegraphics[width=.3\linewidth]{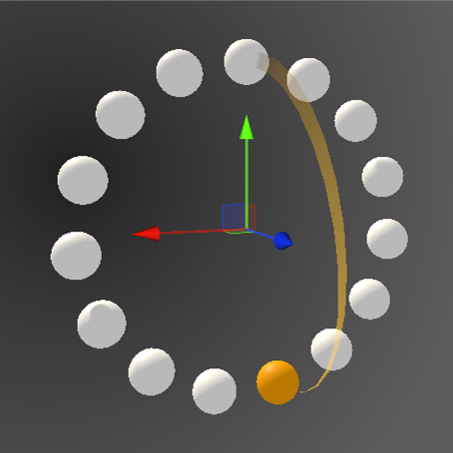}\label{fig:FittsLawTargetsLine}}\;
    \caption{Our Alignment Task targets positions viewed from the front as a layout schematic (a), and the side (b). 15 targets of size 15\textdegree{} visual angle were placed along the blue ring 10m away, where $\theta$ was the radius \{45\textdegree{}, 90\textdegree{}, 135\textdegree{}\}. (c) shows the study setup when $\theta$ is 90\textdegree{}, positioning all the sphere targets around the VR camera (represented as the multicolored axis arrowhead) such that they all lie on the same plane. The orange sphere represents the target to be selected, the white spheres are only shown for figure purposes. The orange arc provides visual feedback connecting the previous target to the next. \label{fig:FittsLawTargetsPositions}}
\end{figure}

Fitts’ law experiments are widely used to test and model interactions that involve efficiently pointing rapidly towards a known and predictable target before selecting it.
Our study draws inspiration from Fitts-style target acquisition tasks to evaluate how effectively users can control the viewport to locate targets in known directions, both within and beyond the initial field of view.
Importantly, the task focuses on bringing targets --- often initially from outside the FOV  --- into central vision rather than rapid target acquisition, distinguishing the task from traditional Fitts’ law paradigms.

In our task, the targets were positioned as shown in \autoref{fig:FittsLawTargetsPositions} and selected via gaze and a button press.
We tested ring radiuses ($\theta$) of 45\textdegree{}, 90\textdegree{}, and 135\textdegree{}, selected to represent a range of head–eye coordination demands.
45\textdegree{} was initially within FOV but not comfortably reachable with the eyes alone; 
90\textdegree{} was initially outside FOV but comfortably reachable via head movements; 
and 135\textdegree{} was both initially outside FOV and also not comfortably reachable via head movements.
Each ring contains a relatively high number of targets (15), providing a higher angular sampling density to capture the fine-grained, directional characteristics of the interaction techniques under investigation.

In the design of traditional Fitts' Law pointing tasks, to ensure the movement is efficient and direct, targets are always kept in view to provide essential visual feedback for corrective (closed-loop) sub-movements to occur during selection ~\cite{McGuffin2005FittsAndExpandingTargets}.
However, the targets in our task have large angular separation ($\theta$) and are mostly not in the FOV.
To maintain visual feedback and help participants locate the targets -- and to discourage “shortcuts” (e.g., rotating in the shorter direction when $\theta> 90^\circ$), a visual guide is displayed connecting the previous and next targets.
The visual guide, shown in \autoref{fig:FittsLawTargetsLine} as the orange arc, can be interpreted as the straight line path traveled between targets in a standard Fitts' Law task, projected onto a sphere centered on the participant to form an arc.
This line is not the shortest arc between targets (i.e., the geodesic), but is part of the circular intersection between a flat plane and the sphere, where the plane intersects the center of both targets, such that the arc spans exactly $\theta$ degrees. 

\subsection{Search Task: Looking towards unknown directions}\label{subsec:evaluation:task2design}
\begin{figure}[tb]
    \centering
    \subfloat{\includegraphics[width=.3\linewidth]{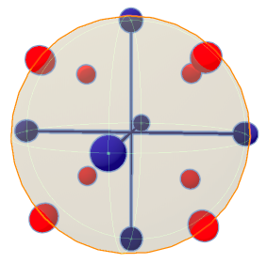}}\;
    \caption{Our Search Task targets positions. 14 targets of size 15\textdegree{} visual angle (2.63m wide) were placed 10m away. Target were placed in directions that can be thought of as extending from the center of a cube to the center of each face (blue) and to each corner (red). Colors and lines are for purely illustrative reasons.\label{fig:SearchTaskTargetsPositions}}
\end{figure}

In our alignment task, the knowledge of target positions guides gaze and viewport control behavior.
However, in real-world scenarios, the locations of points of interest are not necessarily known beforehand, requiring a search phase before interaction.
We therefore adapted a self-directed visual search task to evaluate how effectively users can control the viewport to search and locate targets in unknown directions beyond the initial FOV.
In our task, targets appear one at a time at fixed positions (shown in \autoref{fig:SearchTaskTargetsPositions}), with each new target displayed only after the previous one was selected. 
Positions are selected randomly with two constraints: no position can appear consecutively, and each position appears exactly three times total.

While self-directed visual search beyond the initial FOV in VR has been studied primarily in the context of optimising search strategies and reducing search time~\cite{Stein2024EyeHeadMovementsDuringOutsideFoVSearch, Marek2020CueingBeyondFoV, Grinyer2022FoVEffectsOnOutsideFOVSearch}, our objective differs. 
Rather than focusing on search strategy and efficiency, we wanted to investigate whether exploring all directions can be achieved with minimal effort.
To this end, the target positions (shown in \autoref{fig:SearchTaskTargetsPositions}) were deliberately arranged so participants were required to search in all directions (including directly behind, above and below), thereby ensuring completeness of exploration.

\subsection{Procedure}
\begin{figure}
    \centering
    \subfloat{\includegraphics[height=0.3\linewidth]{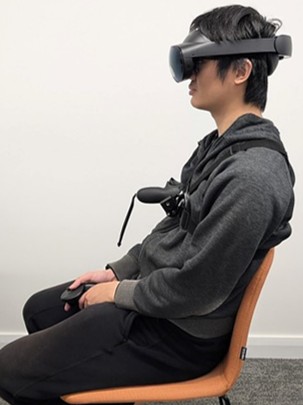}}\;
    \subfloat{\includegraphics[height=0.3\linewidth]{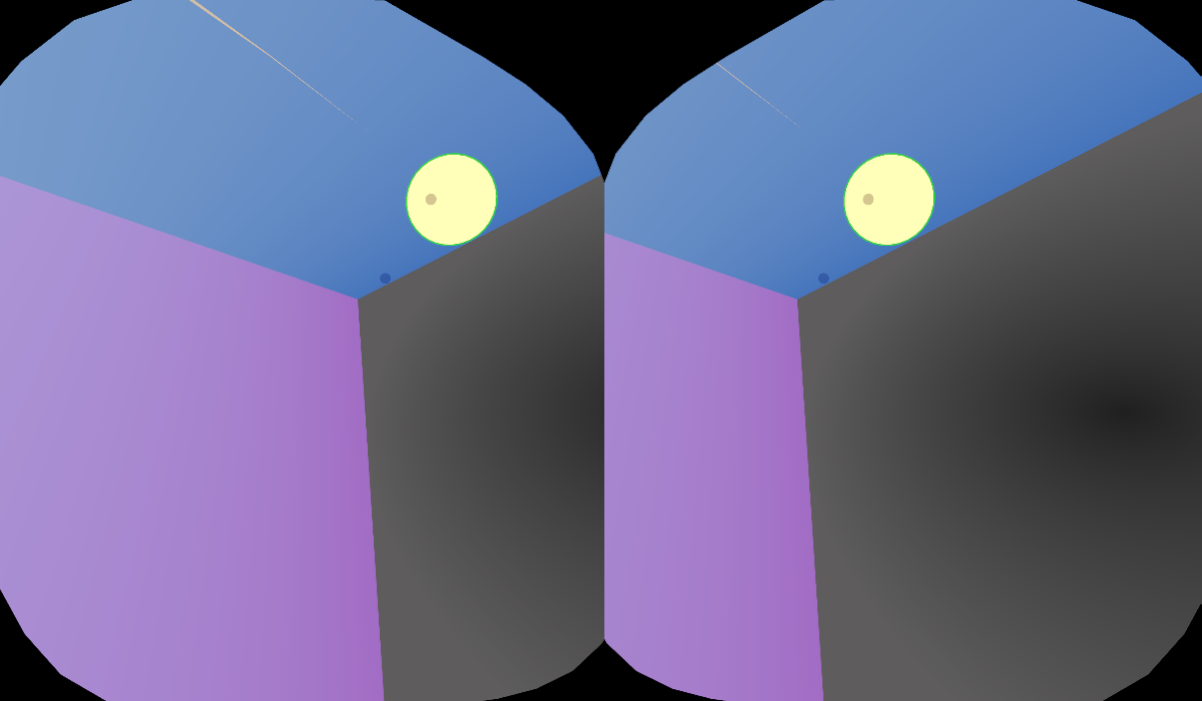}}\;
    \caption{A study participant wearing the chest harness while seated on a stationary chair for the study (left), and what they were seeing in VR (right). The images was taken during the Alignment task when the radius $\theta$ was 135\textdegree{}.}
    \label{fig:StudyParticipant}
\end{figure}
At the start of the task, the experimenter welcomed the participant and explained the purpose and procedure of the task. 
After providing consent, participants filled out a demographics questionnaire asking for age, gender, vision (normal, corrected with glasses, corrected with contact lenses), experience with video games (never, rarely, monthly, weekly, daily), experience with VR (never, rarely, monthly, weekly, daily), and experience with eye tracking (never, rarely, monthly, weekly, daily). 
Then, participants were instructed to put on a provided chest harness with sensors attached for torso tracking (shown in \autoref{fig:StudyParticipant}).
Participants were seated in a stationary, non-rotating chair, and eye-tracking calibration was performed at each headset donning. 
Participants completed both tasks with one technique at a time before moving to the next. Technique order (balanced Latin square) and task order within each technique were counterbalanced across participants.

Before using each technique, participants received a brief explanation of the technique from the experimenter.
Before the start of each task, participants practiced using the technique via a practice task -- a greatly shortened version of the task consisting of 15 selections.
Participants were prompted at the end of each practice task if they wanted to repeat the practice or to continue with the study task.
Most participants chose not to continue practicing, with a rare few choosing to practice an additional time. 
No participants chose to continue practicing more than once.
After completing each task, participants took off the HMD and were given as much time as necessary to rest and complete questionnaires regarding the techniques.
The study in its entirety lasted under 70 minutes.

\subsection{Apparatus}
We implemented and ran the task with Unity 2021.3.27f1 on a computer with an Intel Core i7-12700 CPU, 16 GB RAM, and an NVIDIA GeForce RTX 3070 Ti GPU, connected to a Meta Quest Pro VR HMD with a Quest Link cable. 
The Meta Quest Pro had a 111.24° diagonal FOV, 1800x1920 pixels resolution per eye, and 90 Hz refresh rate.
Torso tracking used a Meta Quest Pro handheld controller attached to the front of a chest harness.

\subsection{Techniques}
The techniques compared in our tasks were:
\begin{itemize}[noitemsep]
    \item 1xHead (baseline) represents the default 1:1 mapping commonly found in VR experiences, where HMD rotation directly maps to viewport rotation.
    \item 2xHead is a HeadGain implementation using our head remapping module (see \autoref{subsec:remappingHeadMovements}) that amplifies head rotation by a factor of 2, doubling the HMD yaw and pitch such that the viewport has twice the amount of yaw and pitch (relative to the forward direction). 
    \item LALA is our gaze-based viewport control technique introduced in \autoref{sec:techniques}.
\end{itemize}

Initially, we considered comparing LALA against straightforward pitch-and-yaw extensions of Gaze Gain and Gaze Pursuit. 
However, oculomotor and perceptual asymmetries (see \autoref{subsec:Techniques:eyeModel} and \autoref{subsec:LALApitch}) prevented these originally 1D yaw-only techniques from being readily extended to combined pitch–yaw rotations. 
Moreover, naive extensions yielded unfair comparisons: pilot studies showed that pitch-and-yaw Gaze Pursuit produced disorienting viewing, while pitch-and-yaw Gaze Gain caused excessive eye strain. 
We also considered unmodified yaw-only Gaze Gain and Gaze Pursuit, but these were unsuitable because our targets vary substantially in pitch, making pitch equivalent to the 1xHead baseline, while yaw performance had already been evaluated in \cite{Lee2024SnapPursuitGainViewportControl}. 
Instead, we included 2xHead as a reference condition because HeadGain techniques are common in prior work, and higher gains such as 2× have not been evaluated for our target scenario.

\subsection{Measures}
Our studies aimed to assess how comfortably and effectively participants could control the viewport to view targets.
For quantitative measures, we recorded the cumulative angular rotation and range of motion of the eyes, head, and torso for each selection. 
We also measured the time taken for the participant’s gaze to first land on the target (i.e., completion time) and the error rate (i.e., missed selections).

For subjective measures, participants rated the following statements on a 7-point Likert scale: “The technique was easy to use.” “The technique was comfortable to use,” and “You had a good sense of control over the technique”.
These ratings provided an overall indication of ease of use, comfort, and perceived controllability.
Participants also completed the RAW NASA-TLX~\cite{Byers1989RawNasaTLX} and VRSQ~\cite{Kim2018VRSQ} questionnaires to capture perceived workload and cybersickness, respectively. 
Finally, participants were invited to rank the techniques according to their preference and share any additional comments.

\subsection{Sample}
We initially recruited 21 participants from the local university. 
However, three were excluded from the final analysis: one participant dropped out immediately due to motion sickness unrelated to the task or technique, another experienced issues with eye-tracking calibration, and the third had an eye injury affecting upward gaze in the right eye. 
This resulted in a final sample of 18 participants (8 identified as male, 10 as female).
The average age was 29.0 years (SD = 8.73; range = 22–53). 
Vision was reported as normal (n = 5), corrected with contact lenses (n = 2), or corrected with glasses (n = 11).
Video gaming frequency was reported as daily (n = 4), weekly (n = 3), monthly (n = 2), less than once a month (n = 7), or never (n = 2).
Both VR usage and eye-tracking usage were reported as none (n = 3), rarely (n = 12), or monthly (n = 3).
All participants received a £10 Amazon voucher. 
The university's Institutional Review Board approved the study.

\section{Results}\label{sec:results}
Quantitative objective data of both tasks were analyzed using two-way repeated measures ANOVA ($\alpha$ =.05). 
When the assumption of sphericity was violated, as indicated by Mauchly’s test, Greenhouse-Geisser corrections were applied (which can cause non-integer degrees of freedom).
Normality was assessed using the Shapiro-Wilk test and Q-Q plots.
If the assumption of normality was violated, the Aligned Rank Transform (ART) procedure~\cite{Wobbrock2011ART} was used.
Post-hoc pairwise comparisons were performed only in cases of statistically significant main or interaction effects and were limited to within-factor comparisons (i.e., conditions differing along a single factor only).
Bonferroni correction was applied where appropriate.
Pairwise $t$-tests were used when parametric assumptions were met, whereas ART-C were applied when assumptions were violated.
Effect sizes are reported as general eta squared ($\eta_g^2$) when parametric assumptions were met, partial eta squared ($\eta_p^2$) when ART was applied.
Subjective data were analyzed using Friedman tests, with effect sizes reported as Kendall’s W.
In all bar charts, only comparisons that are not statistically significant are labeled (ns).
Statistical significance was prioritized in the main text. 
Further statistical details and tables summarizing the statistics for easier viewing are provided in the Appendix.

\subsection{Alignment Task Results}
\begin{table*}[h]
\centering
\resizebox{\linewidth}{!}{%
\begin{tabular}{|cl|lrcl|lrcl|lrcl|}
\hline
\multicolumn{2}{|l|}{} &
  \multicolumn{4}{c|}{Technique $\times$ Radius Interaction Effect} &
  \multicolumn{4}{c|}{Technique Main Effect} &
  \multicolumn{4}{c|}{Radius Main Effect} \\
\multicolumn{1}{|l}{} &
   &
  $F$ value &
  \multicolumn{1}{c}{p} &
  $\eta^2$ &
   &
  $F$ value &
  \multicolumn{1}{c}{p} &
  $\eta^2$ &
   &
  $F$ value &
  \multicolumn{1}{c}{p} &
  $\eta^2$ &
   \\ \hline
\multicolumn{2}{|r|}{Alignment Speed} &
  F(4,136) = 19.532 &
  \textless{}.001 &
  .364 &
  *** &
  F(2,136) = 93.232 &
  \textless{}.001 &
  .578 &
  *** &
  F(2,136) = 332.998 &
  \textless{}.001 &
  .830 &
  *** \\
\multicolumn{2}{|r|}{Error Rate} &
  F(4,136) = .485 &
  .747 &
  .014 &
   &
  F(2,136) = .721 &
  .488 &
  .001 &
   &
  F(2,136) = .773 &
  .463 &
  .011 &
   \\ \hline
\multirow{3}{*}{\begin{tabular}[c]{@{}c@{}}Cumulative\\ Rotation of\end{tabular}} &
  Eye &
  F(1.85,31.47) = 16.153 &
  \textless{}.001 &
  .155 &
  *** &
  F(2,34) = 36.385 &
  \textless{}.001 &   .429 &
  *** &
  F(1.22,20.83) = 256.171 &
  \textless{}.001 &
  .63 &
  *** \\
 &
  Head &
  F(1.73,29.34) = 31.407 &
  \textless{}.001 &
  .241 &
  *** &
  F(1.13,19.21) = 63.309 &
  \textless{}.001 &
  .601 &
  *** &
  F(1.33,22.61) = 346.326 &
  \textless{}.001 &
  .64 &
  *** \\
 &
  Torso &
  F(4,136) = 3.086 &
  \textless{}.018 &
  .083 &
  ** &
  F(2,19.21) = 63.309 &
  \textless{}.001 &
  .601 &
  *** &
  F(2,136) = 36.846 &
  \textless{}.001 &
  .351 &
  *** \\ \hline
\multirow{3}{*}{\begin{tabular}[c]{@{}c@{}}Max\\ Pitch of\end{tabular}} &
  Eye &
  F(4,68) = 23.082 &
  \textless{}.001 &
  .117 &
  *** &
  F(2,34) = 36.385 &
  \textless{}.001 &
  .429 &
  *** &
  F(2,34) = 57.033 &
  \textless{}.001 &
  .187 &
  *** \\
 &
  Head &
  F(1.57,26.69) = 8.995 &
  \textless{}.001 &
  .084 &
  *** &
  F(1.36,23.15) = 83.400 &
  \textless{}.001 &
  .627 &
  *** &
  F(1.47,24.92) = 77.505 &
  \textless{}.001 &
  .258 &
  *** \\
 &
  Torso &
  F(4,136) = 2.605 &
  \textless{}.038 &
  .071 &
  * &
  F(2,136) = 3.015 &
  .052 &
  .050 &
   &
  F(2,136) = 3.527 &
  \textless{}.003 &
  .049 &
  ** \\ \hline
\multirow{3}{*}{\begin{tabular}[c]{@{}c@{}}Max\\ Yaw of\end{tabular}} &
  Eye &
  F(2.69,45.66) = 4.804 &
  \textless{}.001 &
  .039 &
  *** &
  F(2,34) = 34.151 &
  \textless{}.001 &
  .410 &
  *** &
  F(2,34) = 125.892 &
  \textless{}.001 &
  .408 &
  *** \\
 &
  Head &
  F(1.85,31.50) = 34.883 &
  \textless{}.001 &
  .028 &
  *** &
  F(1.18,19.98) = 60.201 &
  \textless{}.001 &
  .606 &
  *** &
  F(1.40,23.81) = 259.584 &
  \textless{}.001 &
  .614 &
  *** \\
 &
  Torso &
  F(4,136) = 10.740 &
  \textless{}.001 &
  .240 &
  *** &
  F(2,136) = 43.026 &
  \textless{}.001 &
  .388 &
  *** &
  F(2,136) = 53.704 &
  \textless{}.001 &
  .441 &
  *** \\ \hline
\end{tabular}%
}
\caption{Alignment Task: ANOVA statistics. Effect sizes reflect $\eta_g^2$ when parametric assumptions were met, and $\eta_p^2$ when ART was applied.}
\label{tab:FittsANOVAs}
\end{table*}

\begin{table}[h]
\resizebox{\linewidth}{!}{%
\small
\begin{tabular}{|cr|lrcl|}
\hline
\multicolumn{2}{|l|}{}           & \multicolumn{4}{c|}{Technique Main Effect}              \\ 
\multicolumn{2}{|l|}{}                           & Friedman Stat & \multicolumn{1}{c}{p} & W    &     \\ \hline

\multirow{2}{*}{\begin{tabular}[c]{@{}c@{}}NASA\\ TLX\end{tabular}} & Physical Demand          & $\chi^2(2)$ = 16.6 & \textless{}.001       & .461 & *** \\
                      & Effort                   & $\chi^2(2)$ = 7.44 & .024                  & .207 & *   \\ \hline
\end{tabular}%
}
\caption{Alignment Task: Significant Friedman statistics for our subjective data. Effect sizes reported as Kendall’s W. Full statistics can be found in the Appendix.}
\label{tab:FittsSubjectiveStats}
\end{table}

\autoref{tab:FittsANOVAs} and \autoref{tab:FittsSubjectiveStats} summarizes the Main and Interaction Effects statistics for our Alignment Task.

\subsubsection{Completion Time}
\begin{figure}[h]
    \centering
    \subfloat[Alignment Task]{\includegraphics[width=.5\linewidth]{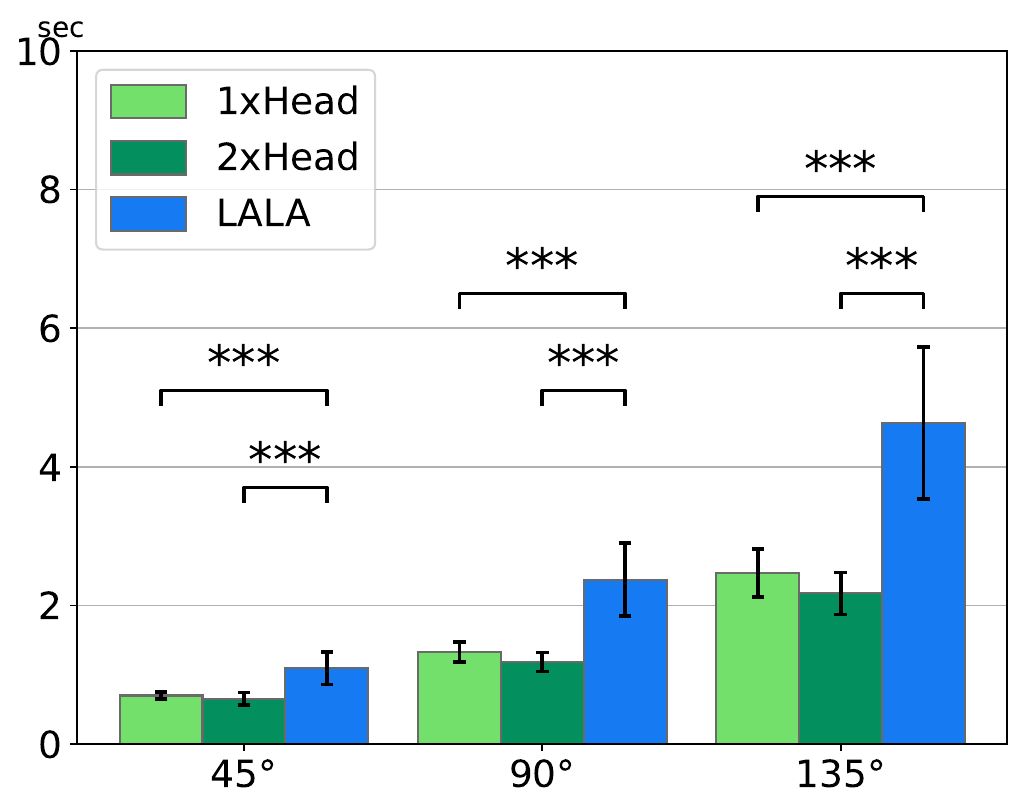}\label{fig:FittsAlignmentSpeed}}
    \subfloat[Search Task]{\includegraphics[height=.27\linewidth*\real{1.5}]{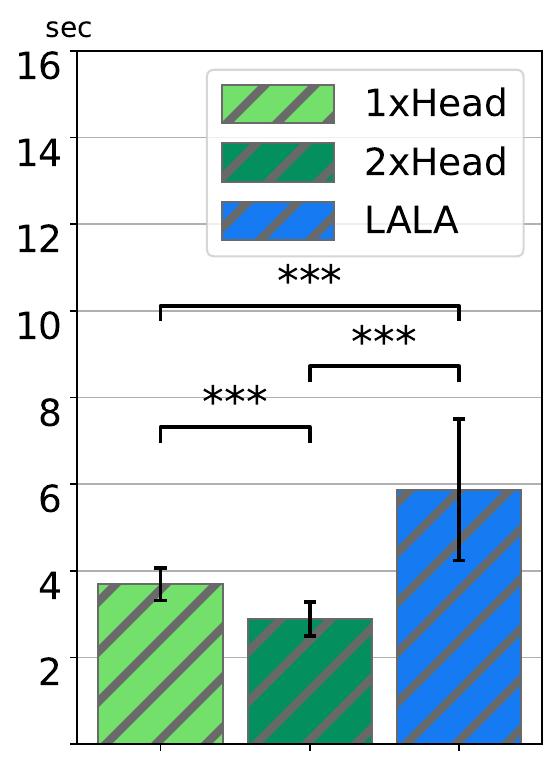}\label{fig:SearchAlignmentSpeedTotal}}
    \caption{Completion Time Per Trial, measured as time taken to gaze at the target, for our Alignment Task by Radius and Technique (a), and for our Search Task (b). Error bars represent the 95\% confidence interval. Lower values are better.}
\end{figure}

There was a significant Technique $\times$ Radius interaction effect, indicating that Technique affected Completion Time (i.e., time taken to gaze at the target) differently at different Radius levels.
Post-hoc pairwise analysis (indicated in \autoref{fig:FittsAlignmentSpeed}) revealed significant differences between all Technique pairs within each Radius level (all p $\leq$ .001) except for the (1xHead, 2xHead) pair (all p $\leq$ 1).
It also revealed significant differences between all Radius pairs within each Technique condition (all p $\leq$ .001).

There were also significant main effects of both Technique and Radius --- though qualified by the significant interaction effect.
Post-hoc pairwise analysis revealed significant differences between all Techniques (all p $<$ .001) and all Radius levels (all p $<$ .001).

\subsubsection{Error Rate}

There were no significant Technique $\times$ Radius interaction effects nor any significant main effects of Technique or Radius, indicating that Error Rate (i.e., missed selections) was similar across all conditions.
The overall mean Error Rate across all conditions was $\mu$ = 0.070 (95\% CI = $\pm$0.015).

\subsubsection{Cumulative Eye, Head, and Torso Rotations}

There was a significant Technique $\times$ Radius interaction effect for Cumulative Eye, Head, and Torso Rotations (see \autoref{tab:FittsANOVAs}), indicating that the effect of Technique varied by Radius level.
Post-hoc pairwise comparisons were conducted between Techniques within each Radius and between Radius levels within each Technique showed that:
For Cumulative Eye Rotation, all Technique pairs differed significantly at each Radius (all p $<$ .019), and all Radius levels differed within each Technique (all p $<$ .001);
For Cumulative Head Rotation, all Technique pairs differed significantly within each Radius (all p $<$ .001) except 2xHead vs LALA (p = 1), and all Radius levels differed within each Technique (all p $<$ .007);
For Cumulative Torso Rotation, no significant differences were found between Techniques or Radius levels (all p = 1).

Significant main effects of Technique and Radius were also found for all three rotation measures.
Post-hoc pairwise analysis revealed that:
For Cumulative Eye Rotation, all Technique and Radius comparisons were significant (all p $<$ .001);
For Cumulative Head Rotation, all were significant (all p $<$ .001) except 2xHead vs LALA (p = .080);
For Cumulative Torso Rotation, all were significant (all p $<$ .001) except 2xHead vs LALA (p = 1).

\subsubsection{Subjective Stats}
\begin{figure}[tb]
    \centering
    \subfloat{\includegraphics[height=.13\linewidth*\real{2}]{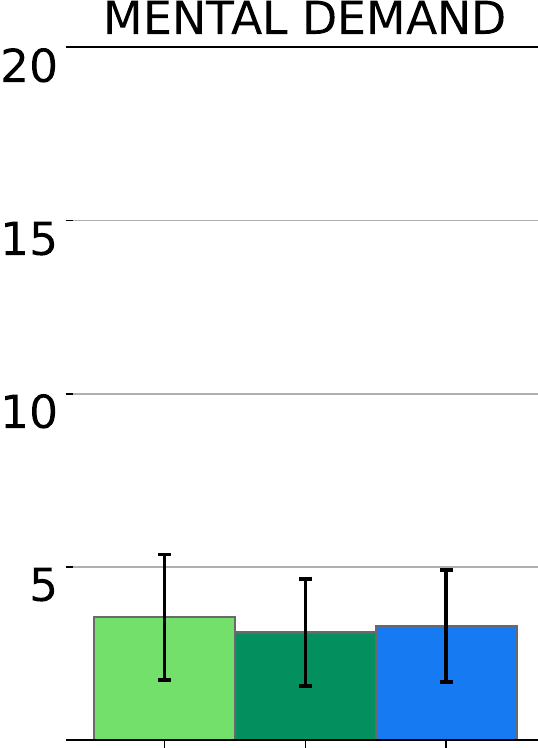 }}
    \subfloat{\includegraphics[height=.13\linewidth*\real{2}]{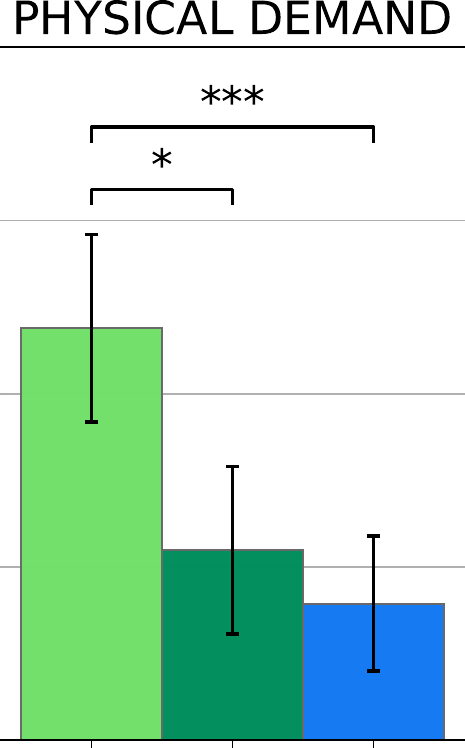}}
    \subfloat{\includegraphics[height=.13\linewidth*\real{2}]{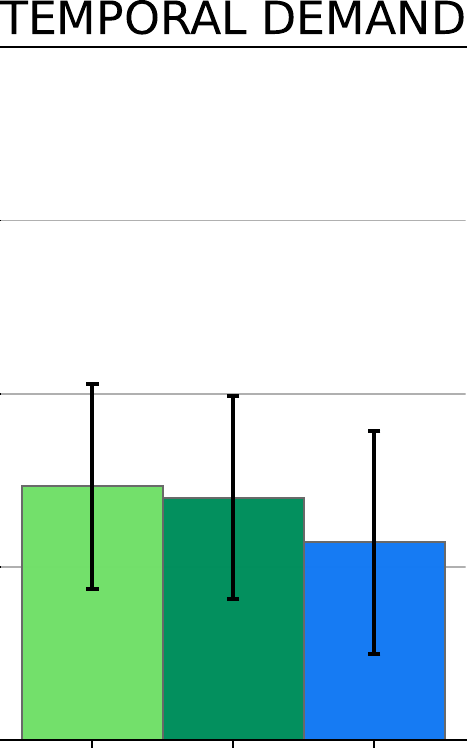}}
    \subfloat{\includegraphics[height=.13\linewidth*\real{2}]{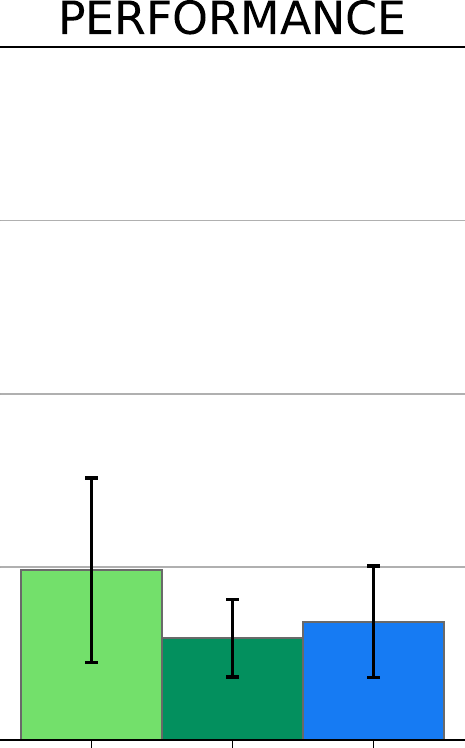}}
    \subfloat{\includegraphics[height=.13\linewidth*\real{2}]{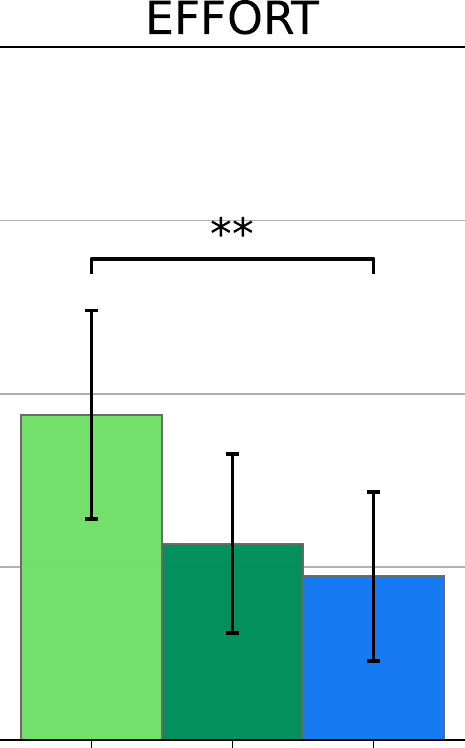}}
    \subfloat{\includegraphics[height=.13\linewidth*\real{2}]{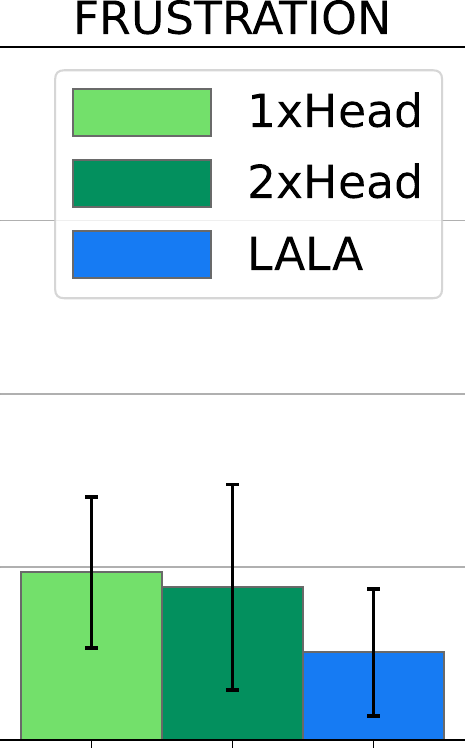}}
    \caption{Alignment Task: Raw NASA TLX by Technique. Error bars represent the 95\% confidence interval. Lower values are better.}
    \label{fig:FittsNASATLXSurvey}
\end{figure}

Analysis of the subjective data collected in our Alignment Task revealed a significant effect of Technique in Physical Demand and Effort, with only non-significant small effects of Technique on other metrics (all p$>$.104). 
Post-hoc Wilcoxon signed-rank tests revealed that 1xHead had significantly higher Physical Demand compared to both 2xHead ($V$=139, p=.010) and LALA ($V$=165, p$<$.001).
Additionally, 1xHead also had significantly higher Effort compared to LALA ($V$=151, p=.008).
There was no significant effect of technique on all VRSQ subscales (all p$>$.683). Overall means($\pm$95\% CI): Oculomotor Disturbance (M = 17.13$\pm$4.82),
Disorientation (M = 11.11$\pm$5.09), and Total Simulator Sickness (M = 14.12$\pm$4.53).

For the remaining subjective metrics, there was no significant effect of Technique (all p$>$.104), with all effect sizes classified as small.
Full subjective data can be found in the Appendix.

\subsection{Search Task Results}
\begin{table}[h]
\resizebox{\linewidth}{!}{%
\small
\begin{tabular}{|cl|lrcl|}
\hline
\multicolumn{2}{|l|}{\multirow{2}{*}{}} & \multicolumn{4}{c|}{Technique Main Effect}                     \\
\multicolumn{2}{|l|}{}                  & $F$ value             & \multicolumn{1}{c}{p} & $\eta^2$ &     \\ \hline
\multicolumn{2}{|r|}{Alignment Speed}   & F(2,34) = 41.43       & \textless{}.001       & .709     & *** \\ \hline
\multirow{2}{*}{\begin{tabular}[c]{@{}c@{}}Cumulative\\ Rotation of\end{tabular}} & Eye & F(1.28,21.73) = 26.422 & \textless{}.001 & .473 & *** \\
                 & Head                 & F(2,34) = 36.58       & \textless{}.001       & .683     & *** \\ \hline
\multirow{2}{*}{\begin{tabular}[c]{@{}c@{}}Max\\ Pitch of\end{tabular}}           & Eye & F(2,34) = 28.942       & \textless{}.001 & .630 & *** \\
                 & Head                 & F(1.2,20.39) = 43.012 & \textless{}.001       & .569     & *** \\ \hline
\multirow{2}{*}{\begin{tabular}[c]{@{}c@{}}Max\\ Yaw of\end{tabular}}             & Eye & F(2,34) = 23.569       & \textless{}.001 & .322 & *** \\
                 & Head                 & F(2,34) = 43.932      & \textless{}.001       & .721     & *** \\  \hline
\end{tabular}
}
\caption{Search Task: Significant ANOVA statistics. Effect sizes reflect $\eta_g^2$ when parametric assumptions were met, and $\eta_p^2$ when ART was applied.}
\label{tab:SearchANOVAs}
\end{table}

\begin{table}[]
\resizebox{\linewidth}{!}{%
\small
\begin{tabular}{|cr|lrcl|}
\hline
\multicolumn{2}{|l|}{}          & \multicolumn{4}{c|}{Technique Main Effect}               \\
\multicolumn{2}{|l|}{}                           & Friedman Stat       & \multicolumn{1}{c}{p} & W    &     \\ \hline
\multirow{4}{*}{\begin{tabular}[c]{@{}c@{}}NASA\\ TLX\end{tabular}}
                      & Physical Demand          & $\chi^2(2)$ = 14.30 & \textless{}.001       & .398 & *** \\
                      & Performance              & $\chi^2(2)$ = 7.30  & .026                  & .203 & *   \\
                      & Effort                   & $\chi^2(2)$ = 8.48  & .014                  & .236 & *   \\
                      & Frustration              & $\chi^2(2)$ = 7.97  & .019                  & .221 & *   \\ \hline
\end{tabular}%
}
\caption{ Search Task: Significant Friedman statistics for subjective data. Effect sizes reported as Kendall’s W. Full statistics can be found in the Appendix.}
\label{tab:SearchSubjectiveStats}
\end{table}

\autoref{tab:SearchANOVAs} and \autoref{tab:SearchSubjectiveStats} summarizes the Main Effects statistics for our Search Task.

\subsubsection{Completion Time}

Analysis of Completion Time, measured as time taken to gaze at the targets, revealed a significant effect of Technique.
As seen in \autoref{fig:SearchAlignmentSpeedTotal}, post-hoc pairwise analysis revealed that 2xHead was significantly faster than 1xHead (\artcyay{12.389}{2.740}{34}{4.521}{.001}) which in turn was significantly faster than LALA (\artcyay{12.556}{2.740}{34}{4.582}{.001}).
Furthermore, 2xHead outperformed LALA (\artcyay{24.944}{2.740}{34}{9.102}{.001}), having the fastest Completion Time overall.

\subsubsection{Error Rate}

Analysis of Error Rate revealed no significant effect of Technique.
The overall mean Error Rate across all conditions was $\mu$ = 0.070 (95\% CI  = $\pm$0.014).

\subsubsection{Subjective Stats}

\begin{figure}[tb]
    \centering
    \subfloat{\includegraphics[height=.13\linewidth*\real{2}]{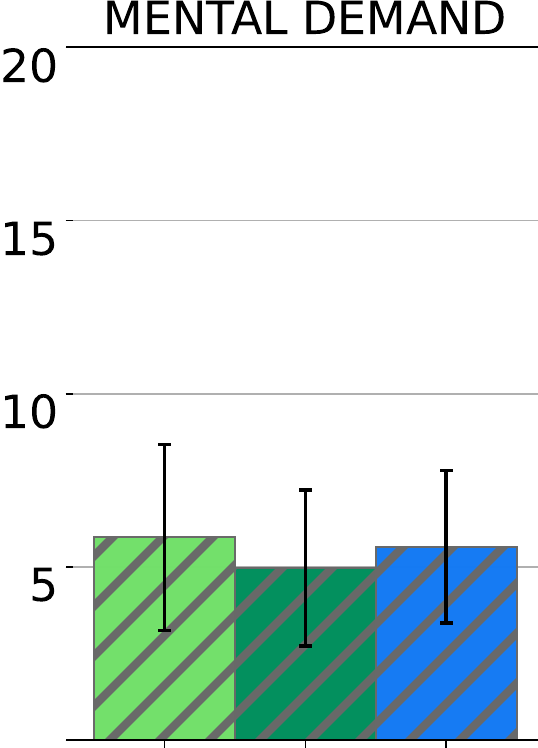 }}
    \subfloat{\includegraphics[height=.13\linewidth*\real{2}]{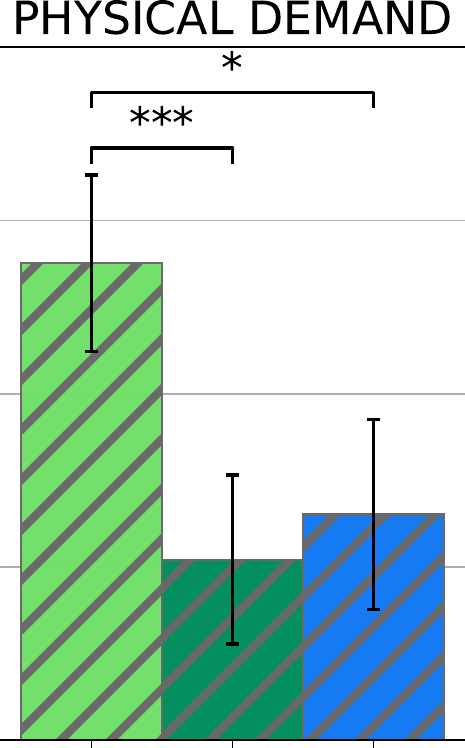}}
    \subfloat{\includegraphics[height=.13\linewidth*\real{2}]{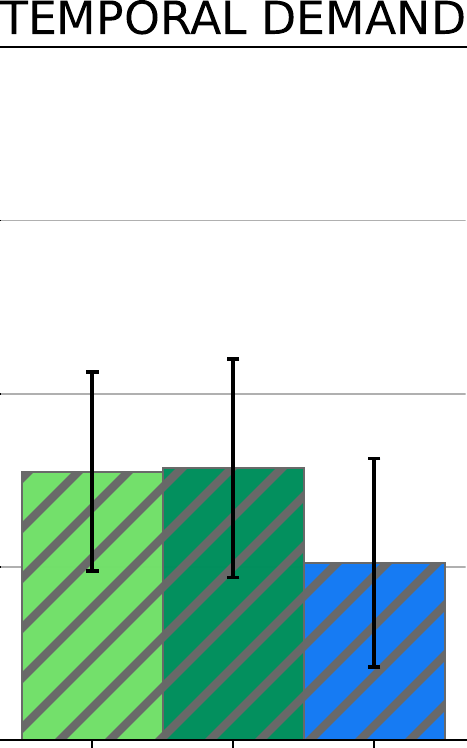}}
    \subfloat{\includegraphics[height=.13\linewidth*\real{2}]{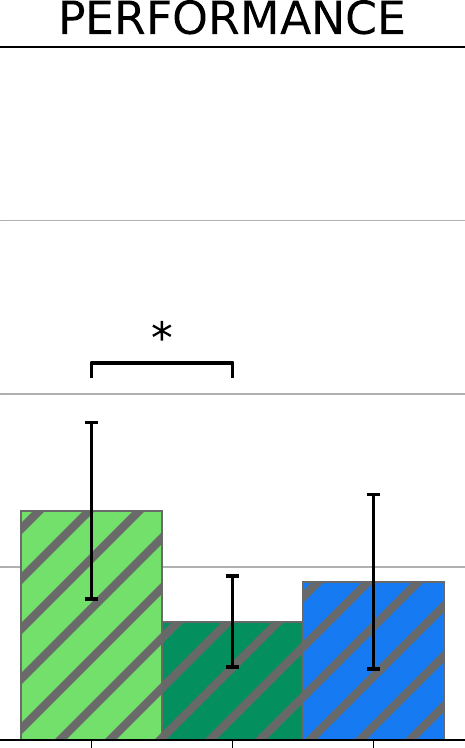}}
    \subfloat{\includegraphics[height=.13\linewidth*\real{2}]{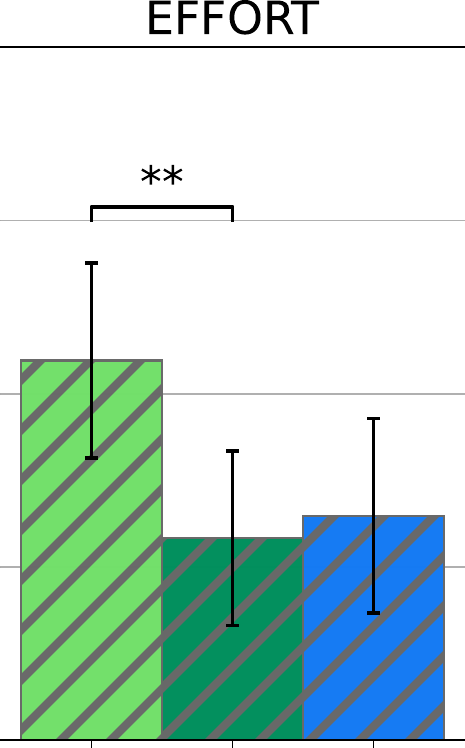}}
    \subfloat{\includegraphics[height=.13\linewidth*\real{2}]{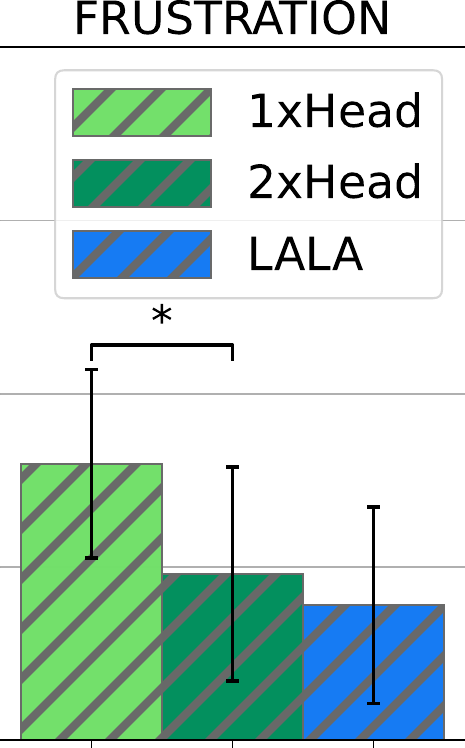}}
    \caption{Search Task: Raw NASA TLX by Technique. Error bars represent the 95\% confidence interval. Lower values are better.}
    \label{fig:SearchNASATLXSurvey}
\end{figure}

Analysis of the subjective data collected in our Search Task revealed a significant effect of Technique in Physical Demand, Performance, Effort, and Frustration.
Post-hoc Wilcoxon signed-rank tests revealed that 1xHead was significantly worse compared to 2xHead in terms of Physical Demand ($V$=165, p$<$.001), Performance ($V$=118, p=.029), Effort ($V$=155, p=.004), and Frustration ($V$=118, p=.032).
Additionally, 1xHead also had significantly worse Physical Demand compared to LALA ($V$=150, p=.002).
There was no significant effect of technique on all VRSQ subscales (all p$>$.059). Overall means($\pm$ CI): Oculomotor Disturbance (M = 23.77$\pm$5.78), Disorientation (M = 16.42$\pm$5.78), and Total Simulator Sickness (M = 20.09$\pm$5.32).

For the remaining subjective metrics, there was no significant effect of Technique (all p$>$.126), with all effect sizes classified as small. 
Full subjective data can be found in the Appendix.

\subsection{User Preference}
\begin{figure}[h]
    \centering
    \includegraphics[width=0.65\linewidth]{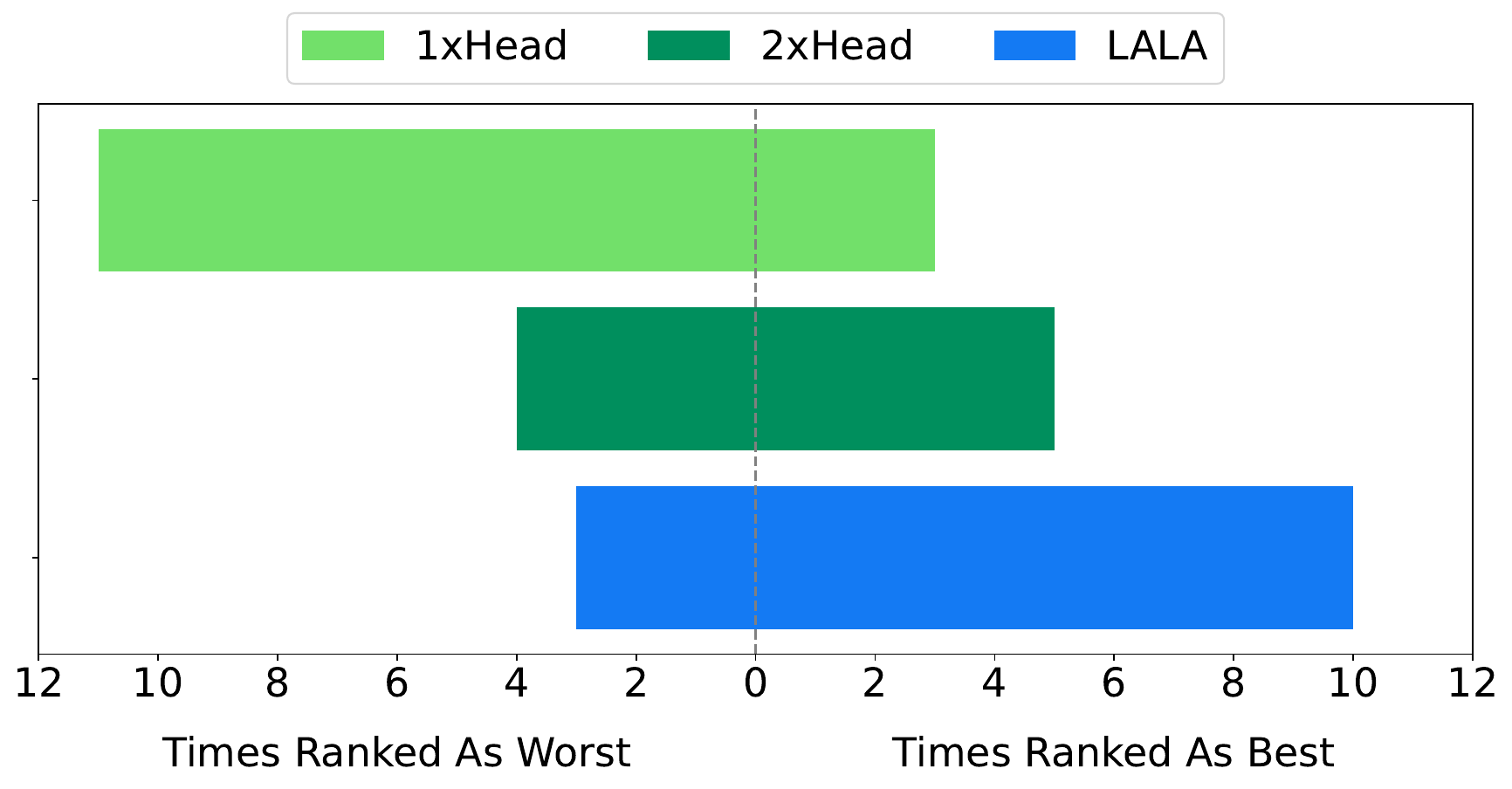}
    \caption{Times each technique was voted as Best or Worst}
    \label{fig:TechniquePreference}
\end{figure}

\autoref{fig:TechniquePreference} presents participants' ratings of the best and worst techniques after experiencing all options.
LALA was rated as the best technique by the most participants (10), followed by 2xHead (5), and 1xHead (3).
Conversely, 1xHead was rated as the worst technique by the most participants (11), followed by 2xHead (4), and LALA (3).
Notably, LALA was rated the best technique by an absolute majority, whereas 1xHead was rated the worst by an absolute majority.

\section{Discussion}\label{sec:discussion}

In this work, we compared three viewport control techniques --- 1xHead, 2xHead, and LALA --- to examine how different control strategies affect performance, effort, and user preferences. 
Across both tasks, LALA was slower but reduced viewing effort compared to the baseline 1xHead and was the most preferred technique overall. 
2xHead was fast but was less preferred, with similar taskload results. 
When directly comparing LALA and 2xHead, participants preferred LALA despite its slower performance.
Both techniques lead to similar levels of effort and error. 
These results highlight a trade-off between the natural 1:1 mapping (1xHead), performance-oriented head amplification (2xHead), and gaze-driven low-effort viewing (LALA).

\subsection{LALA's tradeoff: preference vs speed}
The results reveal a clear performance spectrum across the three techniques.
The baseline 1xHead relies on direct head-to-viewport mapping, providing natural control but requiring large physical rotations that increase physical demand, forcing extreme head movements, ultimately leading to reduced preference.
2xHead amplifies head motion, enabling faster alignment times, particularly for large angular displacements. 
Surprisingly, 2xHead was only faster than 1xHead in the search task, not in the alignment task.
One reason could be that, at extreme angles, the high-control display gain led to overshooting, making it more difficult to perform the alignment.
Combining it with other techniques for precise alignment (e.g., similar to HeadShift \cite{Wang2024HeadShiftHeadPointing}) would alleviate this but break the absolute mapping, introducing complications.
LALA was the slowest. 

LALA being slower than 1xHead and 2xHead was expected, as the technique was intentionally designed and calibrated to prioritize smooth motion over speed.
Scaling up LALA's velocity profiles to a competitive level --- while possible --- would most likely induce cybersickness~\cite{Lo2001CybersicknessSceneRotation, Bonato2009Pitch&RollCybersickness}, lower stability, reduce comfort, and ultimately prevent adoption.
The slower alignment times, therefore, reflect the design trade-off inherent in human-computer interaction: prioritizing users' needs over performance.

Although 2xHead achieved faster task completion times, participants preferred LALA. 
This highlights that objective efficiency alone does not determine user preference.
While 2xHead reduces the amount of head rotation required, the high  control-display gain causes a large visual-vestibular mismatch which can be uncomfortable.
Additionally, it also makes precise selection more challenging.
On the other hand, LALA allows participants to explore the environment primarily through natural eye movements alone.
This leads to a less physical 2D exploration behaviour and is supported by the smoother rotations.
This is supported by the crafted gaze control profile (\autoref{fig:2dEyeModel}) which prevents large accidental rotations while the user is looking around in the central area of the rotated view, only allowing extremely slow rotations that participants seem to not notice or mind.
LALA also has less of a visual-vestibular mismatch compared to 2xHead because only sustained, large, eye-in-head angle causes large rotations.
As sustained large, eye-in-head angles are rare during gaze shifts involving the head, LALA induces minimal extra rotations, being similar to 1xHead.
Still, several participants preferred 1xHead because its direct mapping felt the most natural. 
Together, these findings suggest that viewport control techniques occupy different positions along a trade-off between naturalness, efficiency, and physical effort.

These findings suggest that the three techniques occupy different positions in the design space of viewport control. 
Direct head-based control (1xHead) may remain most suitable for immersive and active VR experiences where natural mapping is important. 
Amplification approaches such as 2xHead may be preferable for tasks that require rapid orientation changes and fast navigation (such as action shooters). 
In contrast, gaze-driven techniques such as LALA may be particularly beneficial for prolonged viewing, slow-paced experiences (escape room games, educational content, 360\textdegree{} videos) or scenarios with constrained mobility (train, airplane, car).

\subsection{LALA's control model for future gaze-based view control}
VR is considered an emerging medium, with adoption still in its infancy~\cite{Richter2023ARVRAdoptionInfographic, Meta2025GoldilocksVRSessionLength}.
LALA illustrates how viewport control can move beyond head rotation alone by exploring alternative interaction strategies based on gaze input, positioning it as an example of a gaze-driven control paradigm rather than a universal successor to head-based techniques.

The results show that our asymmetric omnidirectional gaze control profile (\autoref{subsec:Techniques:eyeModel}) worked well for controlling viewport rotation, allowing participants to select known and search for unknown targets. 
Notably, an additional key benefit is that it encapsulating the complexities of eye movements, reducing the barrier of using eye movements as a control signal, simplifying downstream technique design. 
Prior gaze-based viewport control techniques are often tightly coupled with the visual and oculomotor systems~\cite{Lee2024SnapPursuitGainViewportControl, Lee2024PatternsInMotion}, requiring careful calibration of individual components and making rapid prototyping difficult.
In contrast, because the gaze control profile handles eye behaviour, the calculations of LALA's yaw and pitch velocities are based on relatively simple equations.
We also believe our asymmetric omnidirectional gaze control profile has the potential to be useful for other gaze-based interaction techniques outside viewport control.

A potential extension of LALA is user-specific recalibration of the gaze control profile to account for differences in eye physiology and personal preferences. 
For example, users could roll their eyes to determine their maximum amplitude threshold (blue line in \autoref{fig:2dEyeModel}), which would be incorporated into the eye-effort computation. 
This could also account for individual differences in the tendency to support gaze with head movement (i.e., head movers vs.\ non-head movers)~\cite{Fuller1992Coordination, Thumser2008Coordination}, a factor suggested to underlie issues in prior gaze-based viewport control techniques~\cite{Lee2024SnapPursuitGainViewportControl}.

We believe our design choices position LALA as a promising viewport control technique. Combined with results showing strong user preference, low effort, and high subjective ratings, LALA is a strong candidate for long-term adoption in VR, particularly in accessibility and constrained contexts.
We also expect LALA (and similar techniques) to become more relevant later in VR’s adoption cycle, as the user base broadens to include more diverse physical abilities and as novelty diminishes, increasing demand for more ergonomic, low-effort control methods.

\subsection{Limitations and Future Work}
LALA is the first omnidirectional gaze-based viewport control technique for VR, enabling users to intuitively look around independently of head or torso mobility for general use
However, this study was an abstract evaluation of its feasibility and usability, conducted in a controlled environment to enable consistent comparisons rather than capture the full range of real-world VR use.

Additionally, in our study, participants spent an estimated average of approximately 40 minutes in VR, divided into six segments across an hour with breaks in between.
While this reduced fatigue accumulation and supported controlled comparisons, it also limited fatigue effects that would naturally emerge during longer sessions.\footnote{Meta recommends VR experiences to be designed around a session length of 20 to 40 minutes~\cite{Meta2025GoldilocksVRSessionLength}}
Consequently, the results may not fully generalize to more prolonged, embodied, or immersive VR experiences.

We believe that in highly interactive or immersive applications (e.g., VR games, training, or rehabilitation), naturalness would outweigh ergonomic concerns, favoring the default 1xHead condition. Conversely, in passive, strenuous, or long-duration applications (e.g., 360\textdegree{} videos or virtual workspaces), minimizing effort and ergonomic strain would be prioritized, favoring LALA
Lastly, many interaction techniques use gaze for selection or manipulation. 
We believe LALA can coexist with these approaches due to its natural, implicit interaction style and may even enhance them by reducing required dwell times, as smooth pursuit movements~\cite{Velloso2017MotionCorrelationSelecting, Vidal2013Pursuits} could provide a stronger indicator of user intent, warranting further investigation.

\section{Conclusion}

In this paper, we presented Looking Around by Looking Around (LALA), an omnidirectional gaze-based viewport control technique for VR, designed to minimize physical movement and emphasize intuitiveness. 
We achieved this by first designing asymmetric omnidirectional control profile for the eye, then built on it to exploit tendencies for eyes to stay within comfortable regions for viewport control.
We evaluated LALA in a user study (N=18) featuring two contrasting tasks: alignment towards known directions and open-ended visual search towards unknown directions.
Our results showed that LALA was strongly preferred over the traditional baseline while also enabling fully hands-free interaction with minimal physical movement.

These findings underline that intuitiveness and reduced physical strain can outweigh efficiency in shaping preferences, suggesting that gaze-only viewport control is not only feasible but desirable for many VR contexts. 
Looking ahead, LALA offers gaze-only control for scenarios where comfort and low physical effort are essential, such as extended media viewing, collaborative work, or social VR.
More broadly, LALA reimagines VR as a less physically and spatially demanding medium, expanding its accessibility and usability for prolonged or constrained use.

\acknowledgments{%
This work was supported by the European Research Council (ERC) under the European Union’s Horizon 2020 research and innovation program (Grant No. 101021229 GEMINI: Gaze and Eye Movement in Interaction, and Grant No. 101247164 EyeView360: 360° View Control by Eye Movement for Extended Reality).
}

\bibliographystyle{lib/ISMAR-TVCG-Template/abbrv-doi-hyperref}

\bibliography{bibliography}

\appendix %
\crefalias{section}{appendix} %

\section*{Appendix}\label{sec:appendix}
\subsection{Source Code for LALA and the study}
Source code for LALA and the entire project can be found at
[Redacted for anonymity]

\subsubsection{Pseudo Code for rotation remapping}

\algnewcommand{\LineComment}[1]{\Statex \(\triangleright\) #1}
\begin{algorithm}
\caption{Rotation Remaping}
\begin{algorithmic}[1]
\Procedure{ApplyControllerState}{updatePhase, controllerState}
    \State Execute default XR controller/camera state update
    \State Rotate camera around the local (camera.right.x, 0, camera.right.z) axis by gaze-induced $pitch$.
    \State Rotate camera around the global up axis by gaze-induced $yaw$.
    \item[]
    \LineComment{The following is included for comparison/replication/extensibility and not used in the current experiment}
    \State Rotate xrRig around the local camera right axis by any additional torso-induced $pitch$.
    \State Rotate xrRig around the local camera up axis by any additional torso-induced $yaw$.   
\EndProcedure
\end{algorithmic}
\end{algorithm}

\subsubsection{Representative values for the gaze control profile}

\subsection{Full Statistics}
\begin{table}[h]
\resizebox{\linewidth}{!}{%
\small
\begin{tabular}{|cr|lrcl|}
\hline
\multicolumn{2}{|l|}{}           & \multicolumn{4}{c|}{Technique Main Effect}              \\
\multicolumn{2}{|l|}{}                           & Friedman Statistic & \multicolumn{1}{c}{p} & W    &     \\ \hline
\multicolumn{2}{|r|}{Comfort}                    & $\chi^2(2)$ = 2.98 & .225                  & .082 &     \\
\multicolumn{2}{|r|}{Ease of Use}                & $\chi^2(2)$ = 0.70 & .705                  & .019 &     \\
\multicolumn{2}{|r|}{Controllability}                    & $\chi^2(2)$ = 0.61 & .736                  & .017 &     \\ \hline
\multirow{6}{*}{\begin{tabular}[c]{@{}c@{}}NASA\\ TLX\end{tabular}} & Mental Demand & $\chi^2(2)$ = 0.31 & .985 & .001 &  \\
                      & Physical Demand          & $\chi^2(2)$ = 16.6 & \textless{}.001       & .461 & *** \\
                      & Temporal Demand          & $\chi^2(2)$ = 0.58 & .750                  & .016 &     \\
                      & Performance              & $\chi^2(2)$ = 1.11 & .573                  & .031 &     \\
                      & Effort                   & $\chi^2(2)$ = 7.44 & .024                  & .207 & *   \\
                      & Frustration              & $\chi^2(2)$ = 4.52 & .104                  & .126 &     \\ \hline
\multirow{3}{*}{VRSQ} & Oculomotor Disturbance   & $\chi^2(2)$ = 0.76 & .683                  & .021 &     \\
                      & Disorientation           & $\chi^2(2)$ = 0.54 & .763                  & .015 &     \\
                      & Total Simulator Sickness & $\chi^2(2)$ = 0.03 & .984                  & .001 &     \\ \hline
\end{tabular}%
}
\caption{Allignment Task: Friedman statistics for subjective data. Effect sizes reported as Kendall’s W.}
\label{tab:FittsSubjectiveStatsFull}
\end{table}

\begin{figure}[tb]
    \centering
    \subfloat{\includegraphics[height=.28\linewidth*\real{1.5}]{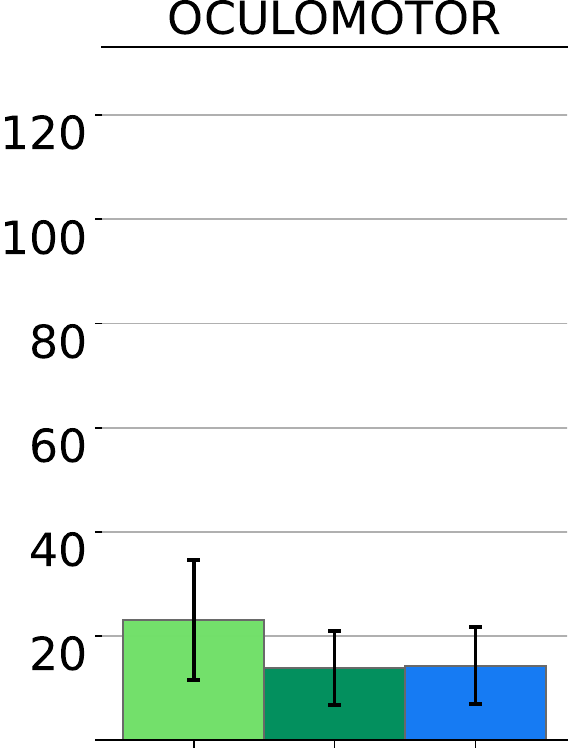 }}
    \subfloat{\includegraphics[height=.28\linewidth*\real{1.5}]{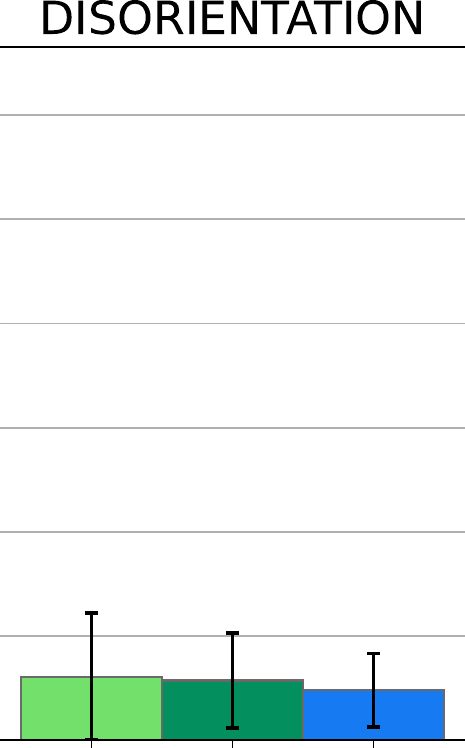}}
    \subfloat{\includegraphics[height=.28\linewidth*\real{1.5}]{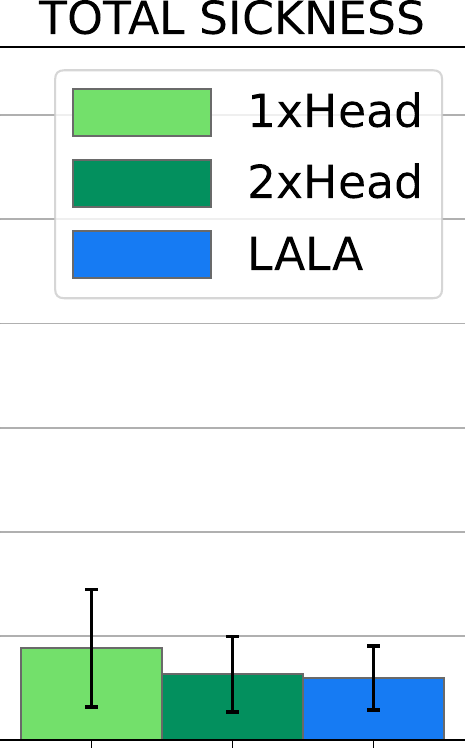}}
    \caption{VRSQ by Technique for our Alignment Task. Error bars represent the 95\% confidence interval. Lower values are better.}
    \label{fig:FittsVRSQSurvey}
\end{figure}
\autoref{tab:FittsSubjectiveStatsFull} and \autoref{fig:FittsVRSQSurvey} shows the remaining full subjective statistics for our Fitts Study.

\begin{table}[h]
\resizebox{\linewidth}{!}{%
\small
\begin{tabular}{|cl|lrcl|}
\hline
\multicolumn{2}{|l|}{\multirow{2}{*}{}} & \multicolumn{4}{c|}{Technique Main Effect}                     \\
\multicolumn{2}{|l|}{}                  & $F$ value             & \multicolumn{1}{c}{p} & $\eta^2$ &     \\ \hline
\multicolumn{2}{|r|}{Alignment Speed}   & F(2,34) = 41.43       & \textless{}.001       & .709     & *** \\
\multicolumn{2}{|r|}{Error Rate}        & F(2,34) = 0.811       & .453                  & .046     &     \\ \hline
\multirow{3}{*}{\begin{tabular}[c]{@{}c@{}}Cumulative\\ Rotation of\end{tabular}} & Eye & F(1.28,21.73) = 26.422 & \textless{}.001 & .473 & *** \\
                 & Head                 & F(2,34) = 36.58       & \textless{}.001       & .683     & *** \\
                 & Torso                & F(2,34) = 0.818       & .450                  & .046     &     \\ \hline
\multirow{3}{*}{\begin{tabular}[c]{@{}c@{}}Max\\ Pitch of\end{tabular}}           & Eye & F(2,34) = 28.942       & \textless{}.001 & .630 & *** \\
                 & Head                 & F(1.2,20.39) = 43.012 & \textless{}.001       & .569     & *** \\
                 & Torso                & F(2,34) = 0.359       & .700                  & .021     &     \\ \hline
\multirow{3}{*}{\begin{tabular}[c]{@{}c@{}}Max\\ Yaw of\end{tabular}}             & Eye & F(2,34) = 23.569       & \textless{}.001 & .322 & *** \\
                 & Head                 & F(2,34) = 43.932      & \textless{}.001       & .721     & *** \\
                 & Torso                & F(2,34) = 0.266       & .768                  & .015     &     \\ \hline
\end{tabular}
}
\caption{Search Task: ANOVA statistics. Effect sizes reflect $\eta_g^2$ when parametric assumptions were met, and $\eta_p^2$ when ART was applied.}
\label{tab:SearchANOVAsFull}
\end{table}

\autoref{tab:SearchANOVAsFull} shows the remaining full quantitative statistics for our Search Study.

\begin{table}[]
\resizebox{\linewidth}{!}{%
\small
\begin{tabular}{|cr|lrcl|}
\hline
\multicolumn{2}{|l|}{}          & \multicolumn{4}{c|}{Technique Main Effect}               \\
\multicolumn{2}{|l|}{}                           & Friedman Stat       & \multicolumn{1}{c}{p} & W    &     \\ \hline
\multicolumn{2}{|r|}{Comfort}                    & $\chi^2(2)$ = 3.97  & .137                  & .110 &      \\
\multicolumn{2}{|r|}{Ease of Use}                & $\chi^2(2)$ = 2.00  & .368                  & .056 &     \\
\multicolumn{2}{|r|}{Controllability}                    & $\chi^2(2)$ = 0.89  & .642                  & .027 &     \\ \hline
\multirow{6}{*}{\begin{tabular}[c]{@{}c@{}}NASA\\ TLX\end{tabular}} & Mental Demand & $\chi^2(2)$ = 0.09 & .958 & .002 &  \\
                      & Physical Demand          & $\chi^2(2)$ = 14.30 & \textless{}.001       & .398 & *** \\
                      & Temporal Demand          & $\chi^2(2)$ = 4.15  & .126                  & .115 &     \\
                      & Performance              & $\chi^2(2)$ = 7.30  & .026                  & .203 & *   \\
                      & Effort                   & $\chi^2(2)$ = 8.48  & .014                  & .236 & *   \\
                      & Frustration              & $\chi^2(2)$ = 7.97  & .019                  & .221 & *   \\ \hline
\multirow{3}{*}{VRSQ} & Oculomotor Disturbance   & $\chi^2(2)$ = 5.67  & .059                  & .158 &     \\
                      & Disorientation           & $\chi^2(2)$ = 1.65  & .437                  & .046 &     \\
                      & Total Simulator Sickness & $\chi^2(2)$ = 2.84  & .241                  & .079 &     \\ \hline
\end{tabular}%
}
\caption{ Search Task: Friedman statistics for subjective data. Effect sizes reported as Kendall’s W.}
\label{tab:SearchSubjectiveStatsFull}
\end{table}

\begin{figure}[tb]
    \centering
    \subfloat{\includegraphics[height=.28\linewidth*\real{1.5}]{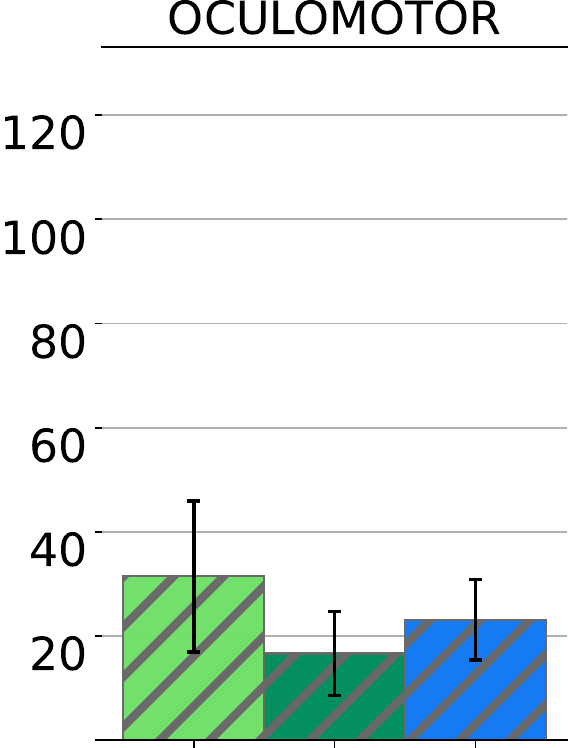 }}
    \subfloat{\includegraphics[height=.28\linewidth*\real{1.5}]{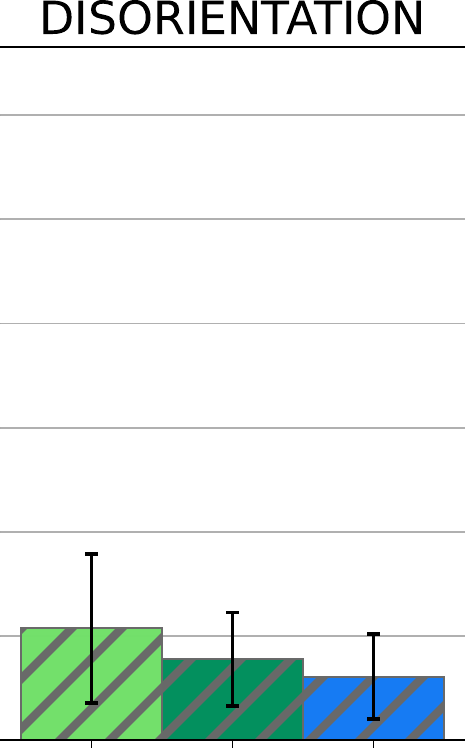}}
    \subfloat{\includegraphics[height=.28\linewidth*\real{1.5}]{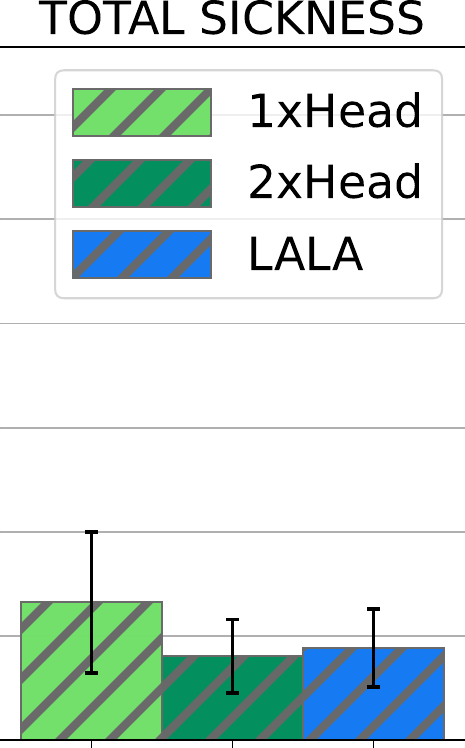}}
    \caption{VRSQ by Technique for our Search Task. Error bars represent the 95\% confidence interval. Lower values are better.}
    \label{fig:SearchVRSQSurvey}
\end{figure}

\autoref{tab:SearchSubjectiveStatsFull} and \autoref{fig:SearchVRSQSurvey} shows the remaining full subjective statistics for our Search Study.

\subsection{Search Study Results by Direction}\label{sec:SearchResultsByDirection}
\begin{table}[h]
\resizebox{\linewidth}{!}{%
\small
\begin{tabular}{|cl|lrcl|}
\hline
\multicolumn{2}{|l|}{\multirow{2}{*}{}} &
  \multicolumn{4}{c|}{Technique $\times$ Direction Interaction Effect} \\
\multicolumn{2}{|l|}{}           & $F$ value          & \multicolumn{1}{c}{p} & $\eta^2$ &     \\ \hline
\multicolumn{2}{|r|}{Alignment Speed} &
  F(26,697) = 5.757 &
  \textless{}.001 &
  .177 &
  *** \\
\multicolumn{2}{|r|}{Error Rate} & F(26,697) = 1.724  & .014                  & .060     & *   \\ \hline
\multirow{3}{*}{\begin{tabular}[c]{@{}c@{}}Cumulative\\ Rotation of\end{tabular}} &
  Eye &
  F(26,697) = 5.245 &
  \textless{}.001 &
  .164 &
  *** \\
             & Head              & F(26,697) = 10.521 & \textless{}.001       & .282     & *** \\
             & Torso             & F(26,697) = 1.79   & .009                  & .063     & **  \\ \hline
\multirow{3}{*}{\begin{tabular}[c]{@{}c@{}}Max\\ Pitch of\end{tabular}} &
  Eye &
  F(26,442) = 2.764 &
  \textless{}.001 &
  .049 &
  *** \\
             & Head              & F(26,442) = 3.783  & \textless{}.001       & .064     & *** \\
             & Torso             & F(26,697) = 1.275  & .164                  & .045     &     \\ \hline
\multirow{3}{*}{\begin{tabular}[c]{@{}c@{}}Max\\ Yaw of\end{tabular}} &
  Eye &
  F(26,442) = 3.074 &
  \textless{}.001 &
  .052 &
  *** \\
             & Head              & F(26,697) = 3.874  & \textless{}.001       & .126     & *** \\
             & Torso             & F(26,697) = 4.201  & \textless{}.001       & .135     & *** \\ \hline
\end{tabular}
}
\caption{Search Task by Direction: ANOVA statistics. Effect sizes ($\eta^2$) reflect $\eta_g^2$ when parametric assumptions were met, and $\eta_p^2$ when ART was applied. Non-integer degrees of freedom are due to Greenhouse–Geisser corrections.}
\label{tab:SearchANOVAsByDirection}
\end{table}

Further analysis was done on the search study results based on the direction of the targets.
The analysis showed that the results for LALA and 2xHead were largely direction-independent, whereas 1xHead exhibited significant performance degradation in the rear hemisphere.

\subsubsection{Completion Time}
\begin{figure}[h]
    \centering
    \subfloat{\includegraphics[height=.42\linewidth]{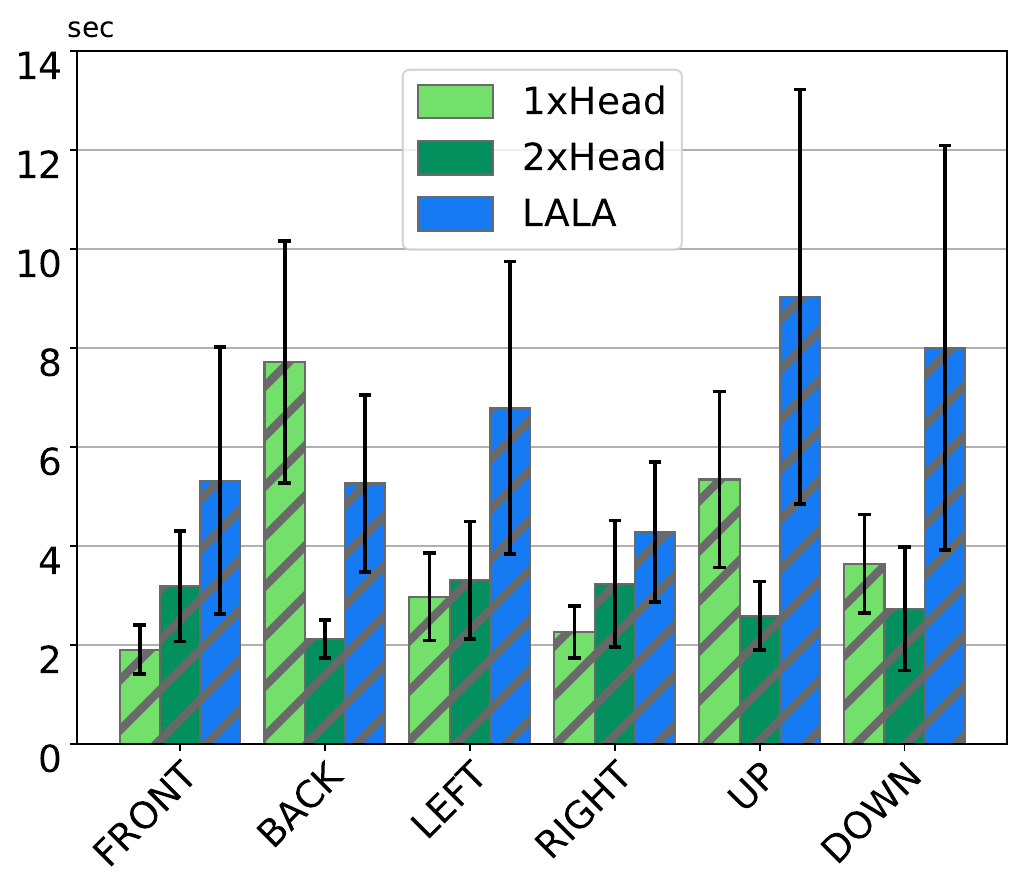}\label{fig:SearchAlignmentSpeedByDirection}}
    \caption{Completion Time By Direction -  only cardinal directions are shown for conciseness. Error bars represent the 95\% confidence interval. Lower values are better.}
\end{figure}
Completion Time broken down by direction (see \autoref{fig:SearchAlignmentSpeedByDirection}) revealed a significant Technique $\times$ Direction interaction effect (see \autoref{tab:SearchANOVAsByDirection}), indicating that Technique affected Completion Time differently at different Directions.
Post-hoc pairwise analysis revealed that while LALA and 2xHead had no within-technique differences between Directions (all p = 1), 1xHead had significantly slower alignment times for targets in the Back Direction, compared to those to the sides or in the frontal hemisphere (all p $<$ .021).
Meanwhile, in the Back and Up Directions, 2xHead was significantly faster than both 1xHead and LALA (all p $<$ .002), while in the Front direction, 1xHead outperformed LALA (p = 0.02).

\subsubsection{Error Rate}
Error Rate broken down by Direction revealed a significant Technique $\times$ Direction interaction (see \autoref{tab:SearchANOVAsByDirection}).
Post-hoc pairwise comparisons, however, showed no significant differences between individual conditions, indicating that Error Rates were similar across the board.

\subsubsection{Max Eye, Head, and Torso Rotations}
\begin{figure}[h]
    \centering
    \subfloat[Max Eye Pitch]{\includegraphics[width=.5\linewidth]{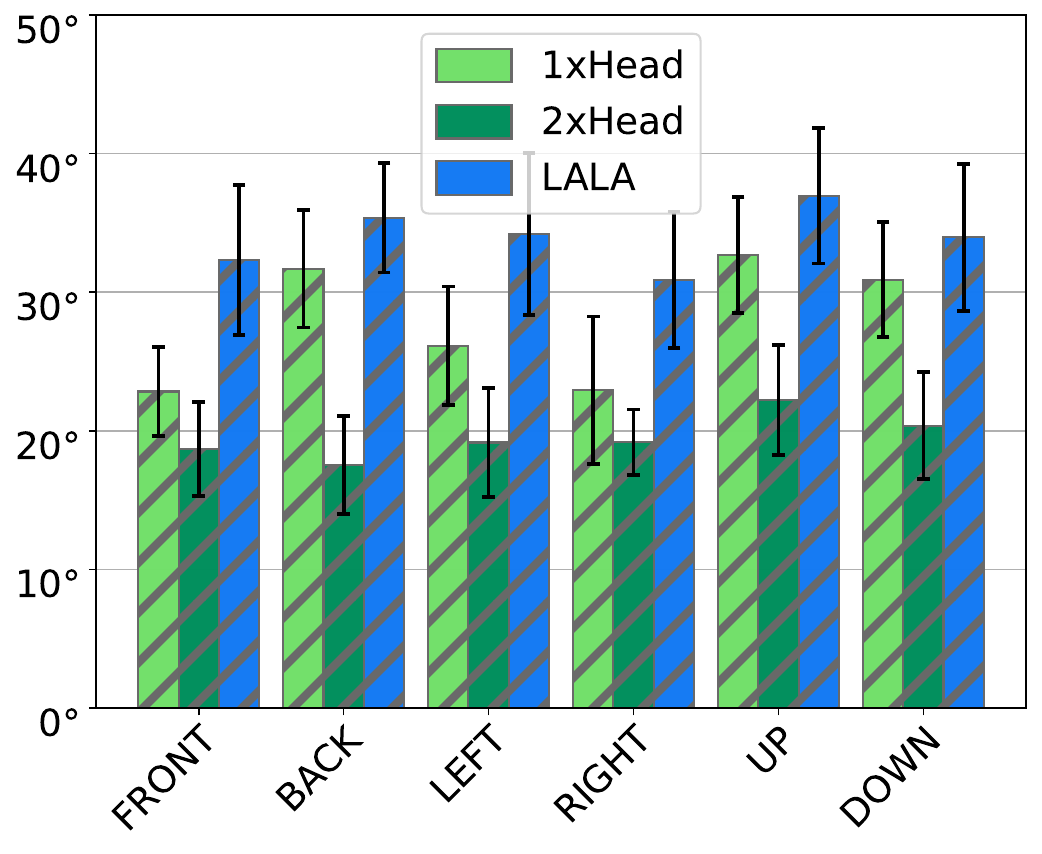}}
    \subfloat[Max Eye Yaw]{\includegraphics[width=.5\linewidth]{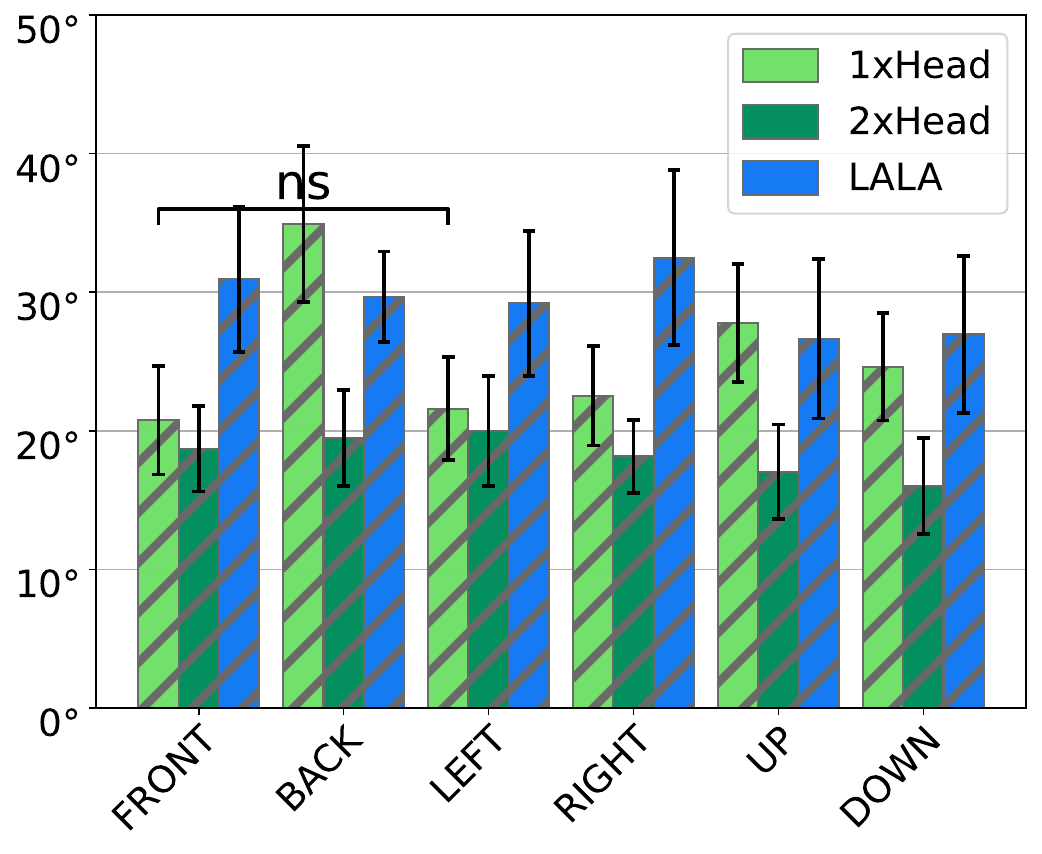}}\\
    \hspace{\linewidth*\real{0.02}}
    \subfloat[Max Head Pitch]{\includegraphics[width=.5\linewidth]{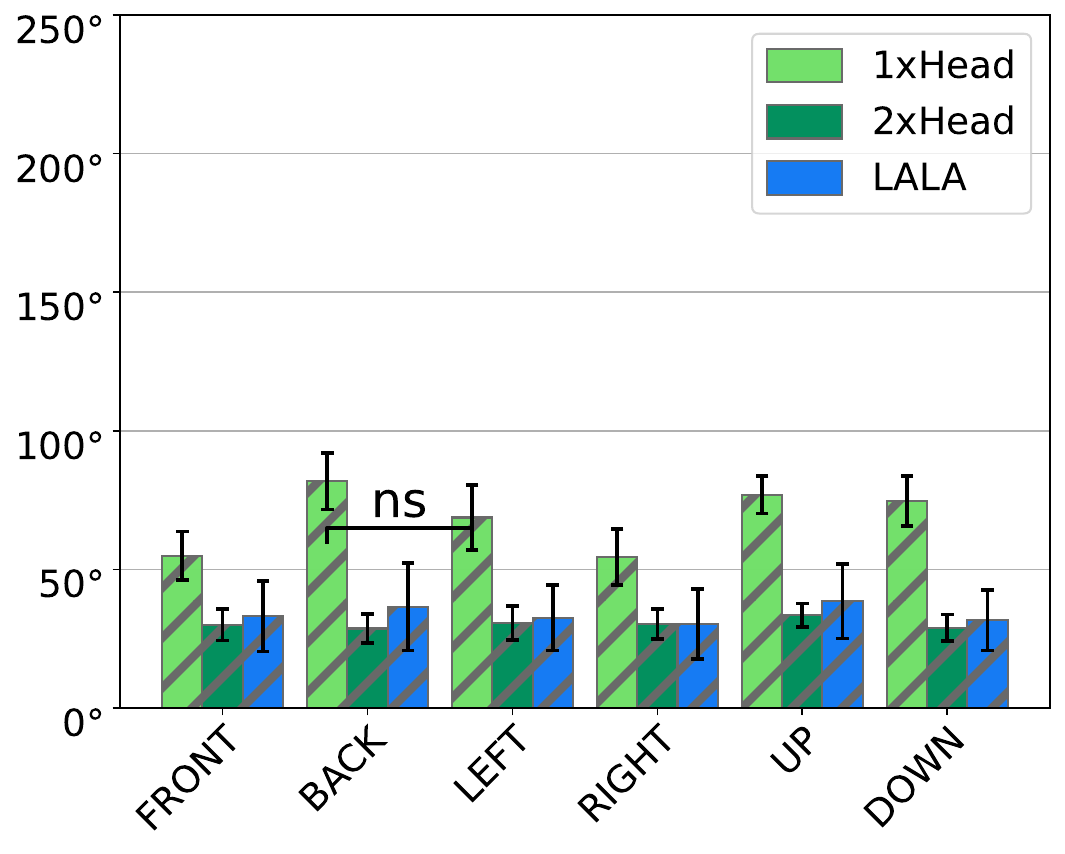}}
    \subfloat[Max Head Yaw]{\includegraphics[width=.5\linewidth]{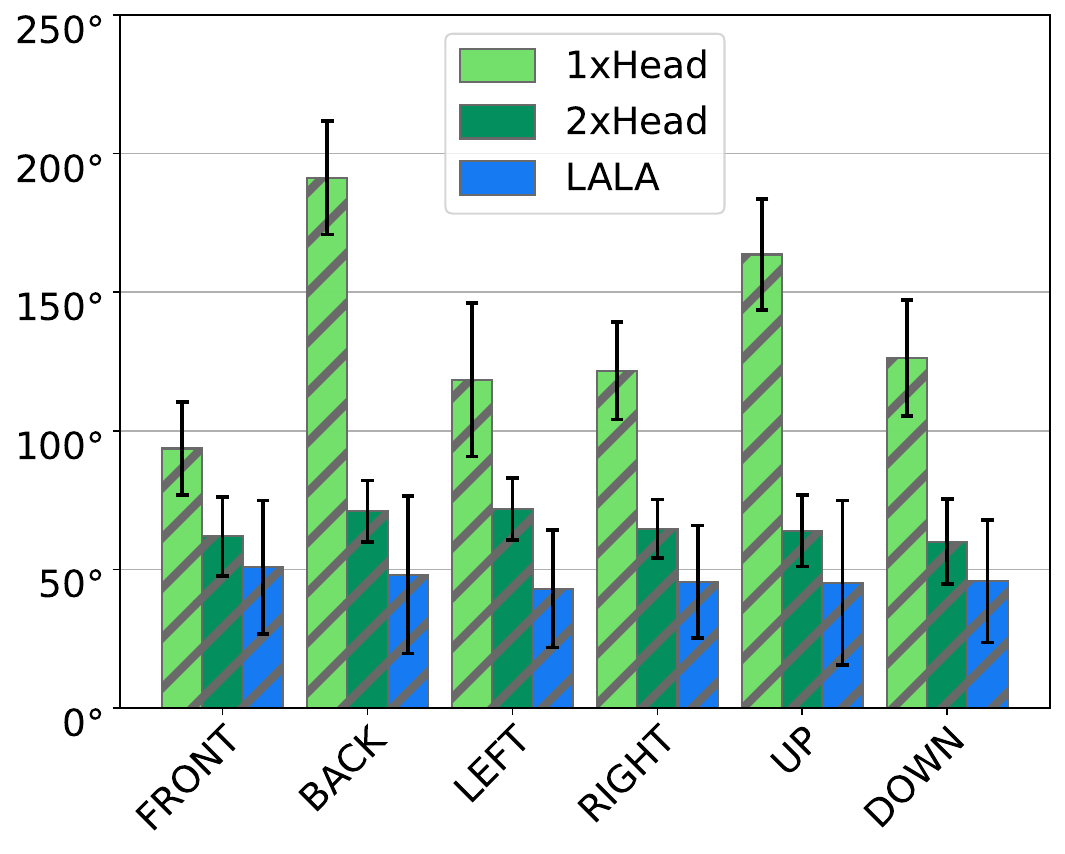}}\\
    \hspace{\linewidth*\real{0.02}}
    \subfloat[Max Torso Pitch]{\includegraphics[width=.5\linewidth]{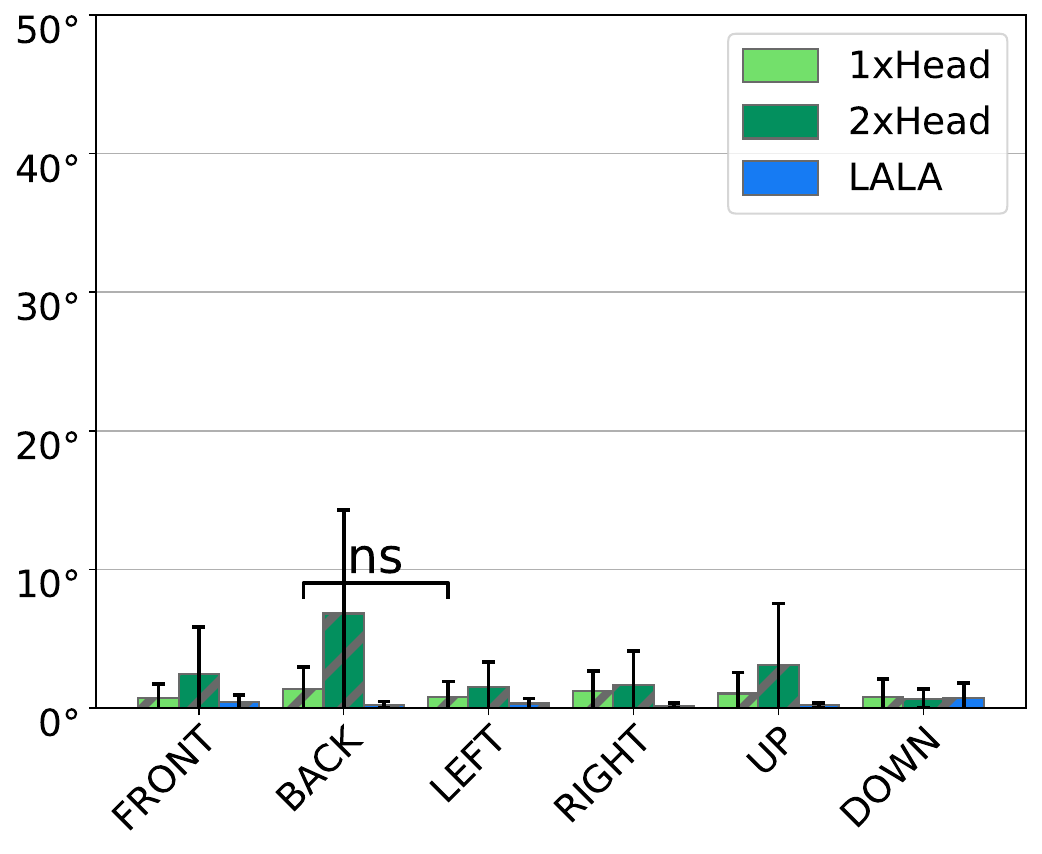}}
    \subfloat[Max Torso Yaw]{\includegraphics[width=.5\linewidth]{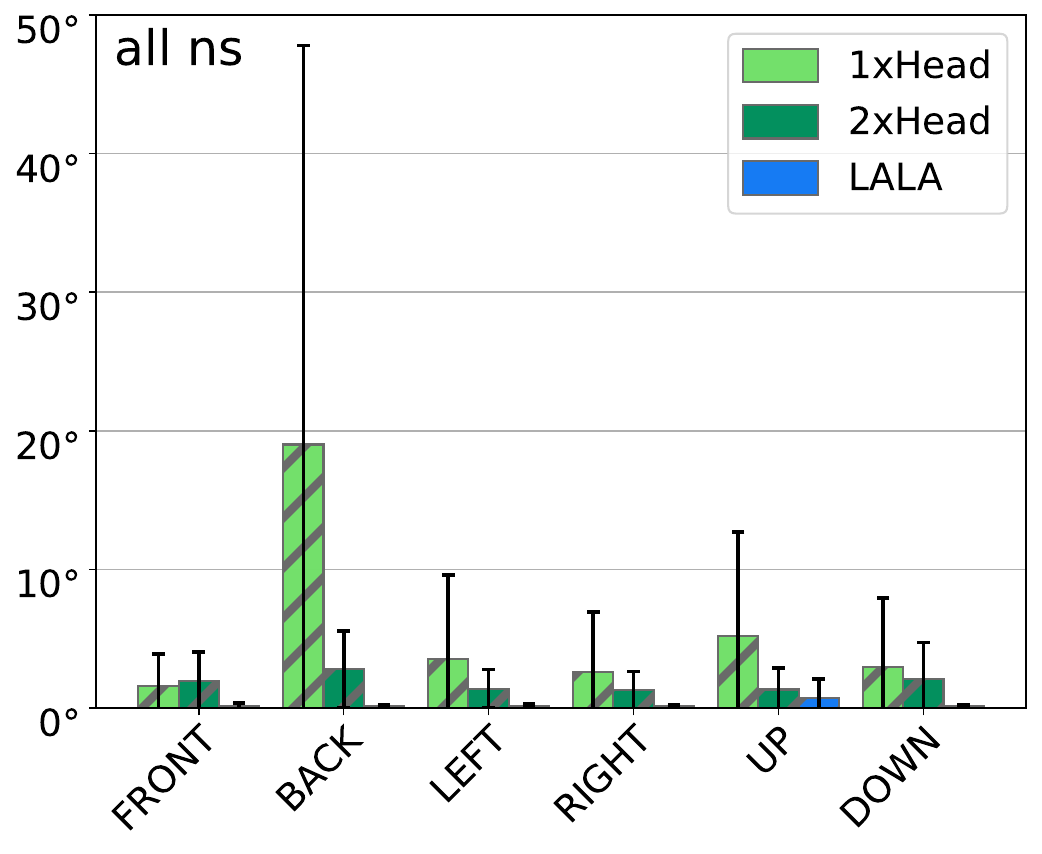}}
    \caption{Max Eye (a), Head (c), and Torso Pitch (e) by Direction and Max Eye (b), Head (d), and  Torso Yaw (f) by Direction - only cardinal directions are shown for conciseness. Error bars represent the 95\% confidence interval. Lower values are better.}
    \label{fig:SearchMaxMovementStatsByDirection}
\end{figure}

Analysis of Max Eye Pitch broken down by Direction (see \autoref{fig:SearchCumulativeEyeByDirection}) revealed a significant Technique $\times$ Direction interaction effect (see \autoref{tab:SearchANOVAsByDirection}), indicating that Technique affected Max Eye Pitch differently at different Directions.
Post-hoc within-technique analysis revealed that only 2xHead had no significant differences in Max Eye Pitch across directions (all p = 1).

Analysis of Max Eye Yaw broken down by Direction (see \autoref{fig:SearchCumulativeEyeByDirection}) revealed a significant Technique $\times$ Direction interaction effect (see \autoref{tab:SearchANOVAsByDirection}), indicating that Technique affected Max Eye Yaw differently at different Directions.
Post-hoc within-technique analysis revealed that while both LALA and 2xHead had no significant differences in Max Eye Pitch across directions (all p = 1), 1xHead had significantly higher Max Eye Yaw for Back compared to all directions in the frontal hemisphere and sides (all p $<$ .039).

Analysis of Max Head Pitch broken down by Direction (see \autoref{fig:SearchMaxMovementStatsByDirection}) revealed a significant Technique $\times$ Direction interaction effect (see \autoref{tab:SearchANOVAsByDirection}), indicating that Technique affected Max Head Pitch differently at different Directions.
Post-hoc analysis revealed that both LALA and 2xHead had no significant differences in Max Head Pitch across directions (all p = 1), and had significantly lower Max Head Pitch than 1xHead in all Directions.

Analysis of Max Head Yaw broken down by Direction (see \autoref{fig:SearchMaxMovementStatsByDirection}) revealed a significant Technique $\times$ Direction interaction effect (see \autoref{tab:SearchANOVAsByDirection}), indicating that Technique affected Max Head Yaw differently at different Directions.
Post-hoc analysis revealed that both LALA and 2xHead had no significant differences in Max Head Pitch across directions (all p = 1).
Meanwhile, 1xHead had significantly higher Max Head Yaw at the Back compared to the Front (p $<$ .001), had significantly higher Max Head Yaw in all directions compared to LALA (all p $<$ .006), and for all directions in the rear hemisphere plus sides compared to 2xHead (all p $<$ .002).

Analysis of Max Torso Pitch by Direction showed no significant Technique $\times$ Direction interaction.
However, there was a significant interaction for Max Torso Yaw by Direction(see \autoref{tab:SearchANOVAsByDirection}).
Post-hoc pairwise comparisons for Max Torso Yaw revealed significant differences only for 1xHead, with the Back direction showing higher values than all other directions.

\subsubsection{Cumulative Eye, Head, and Torso Rotations}
\begin{figure}[h]
    \centering
    \subfloat[Cumulative Eye Rotation by Direction]{\includegraphics[height=.32\linewidth*\real{.8}]{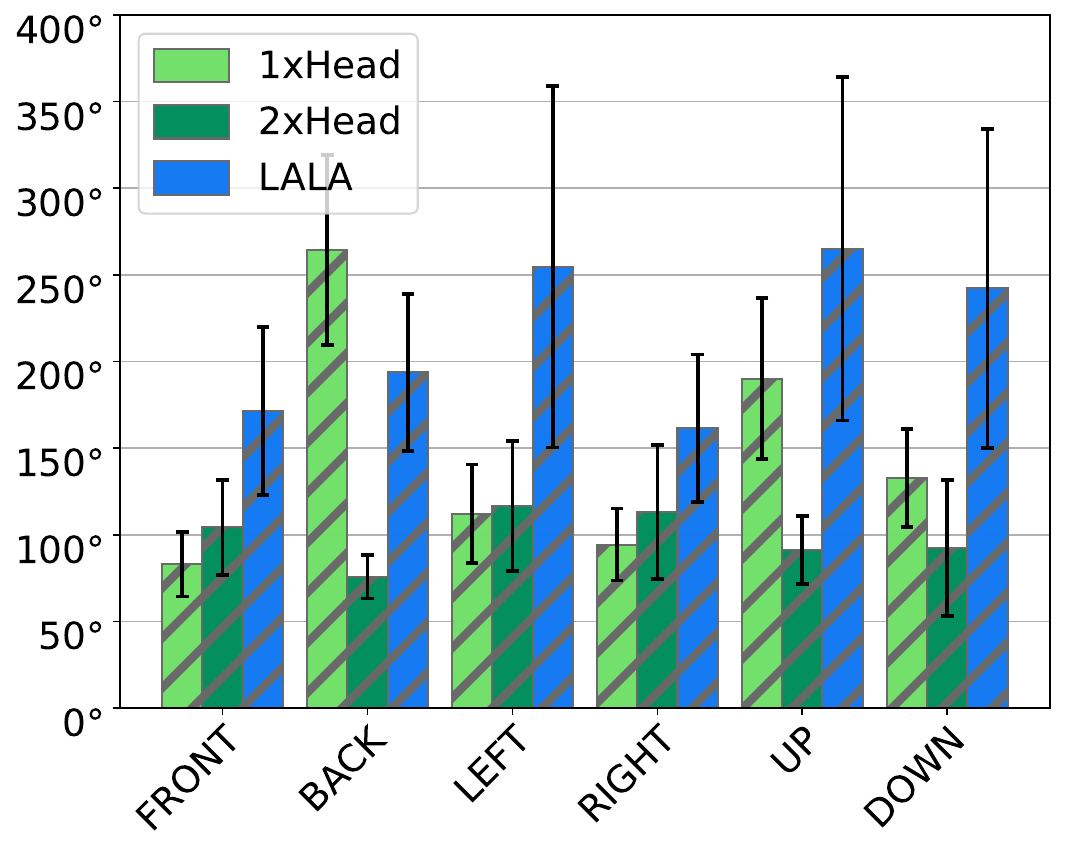}\label{fig:SearchCumulativeEyeByDirection}}
    \subfloat[Cumulative Head Rotation by Direction]{\includegraphics[height=.32\linewidth*\real{.8}]{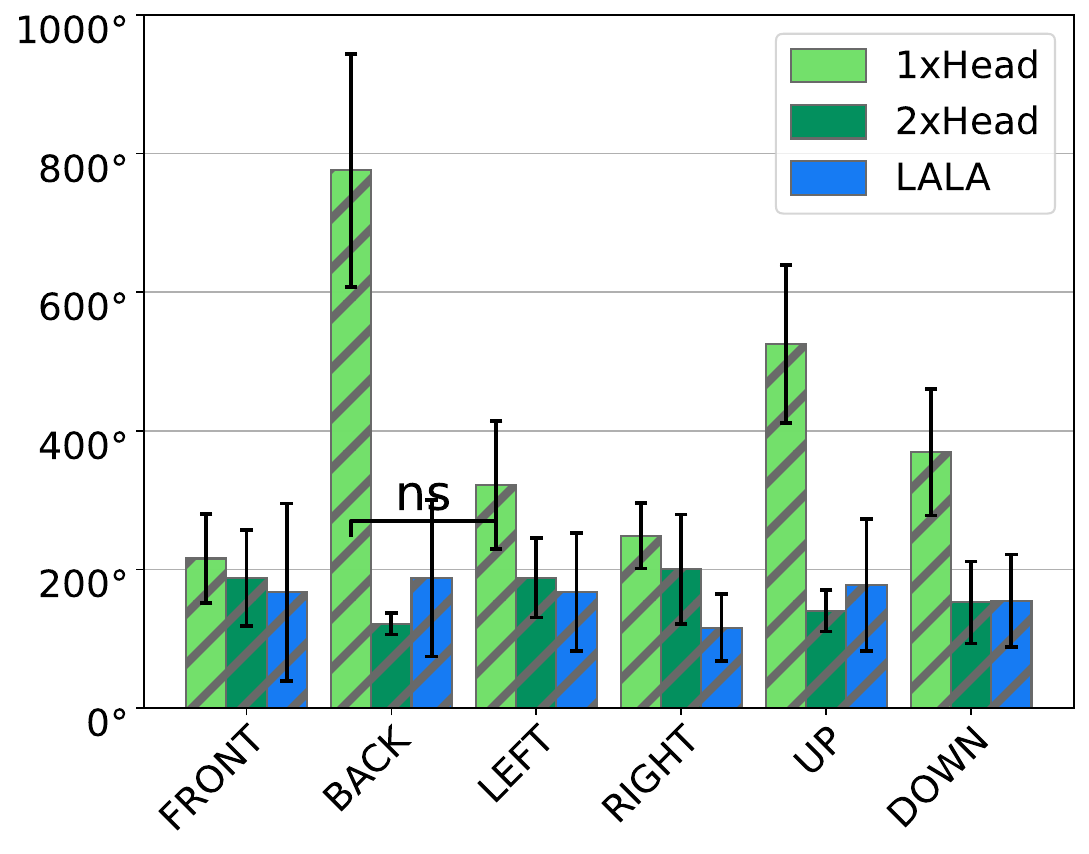}\label{fig:SearchCumulativeHeadByDirection}}
    \subfloat[Cumulative Torso Rotation by Direction]{\includegraphics[height=.32\linewidth*\real{.8}]{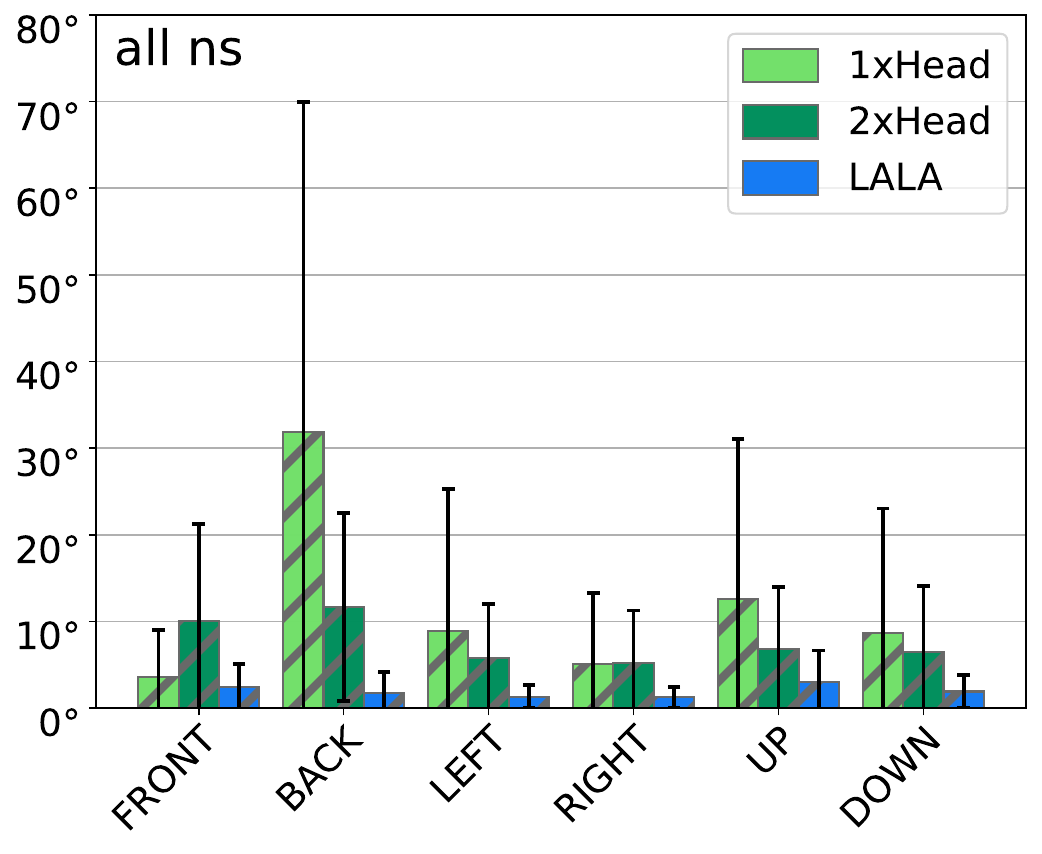}}
    \caption{Cumulative (a) Eye, (b) Head, and (c) Torso Rotation by Direction - only cardinal directions are shown for conciseness. Error bars represent the 95\% confidence interval. Lower values are better.}
    \label{fig:SearchCumulativeMovementStats}
\end{figure}

Cumulative Eye Rotation broken down by direction (see \autoref{fig:SearchCumulativeEyeByDirection}) revealed a significant Technique $\times$ Direction interaction effect (see \autoref{tab:SearchANOVAsByDirection}), indicating that Technique affected Cumulative Eye Rotation differently at different Directions.
Post-hoc within-technique pairwise analysis revealed that while LALA and 2xHead had no significant differences between Directions (all p = 1), 1xHead had significantly more Cumulative Eye Rotation for targets in the upper-rear semihemisphere --- including the Back direction --- compared to those to the sides or in the frontal hemisphere (all p $<$ .032).
Several between-technique, within-direction differences were also observed.
Notably, 1xHead showed less Cumulative Eye Rotation than LALA for the Front direction (p=0.18), while 2xHead had significantly lower rotation than both 1xHead and LALA in the Up direction and most of the directions in the rear hemisphere (all p $<$ .017).
Additionally, both 1xHead and 2xHead had significantly less Cumulative Eye Rotation than LALA for the Front-Up-Right direction (all p $<$ .011).

Cumulative Head Rotation broken down by direction (see \autoref{fig:SearchCumulativeHeadByDirection}) revealed a significant Technique $\times$ Direction interaction effect (see \autoref{tab:SearchANOVAsByDirection}), indicating that Technique affected Cumulative Head Rotation differently at different Directions.
Post-hoc within-technique pairwise analysis revealed that while LALA and 2xHead had no significant differences between Directions (all p = 1), 1xHead had significantly more Cumulative Head Rotation for targets in the upper-rear semihemisphere --- including the Back direction --- compared to those to the sides or in the frontal hemisphere (all p $<$ .017).
Several between-technique, within-direction differences were also observed.
When compared to LALA, 1xHead showed consistently higher head rotation across all directions (all p $<$ .012) except Front-Up-Right (p = 1). 
Additionally, for the Up, Down, and all directions in the rear hemisphere, 1xHead produced significantly more Cumulative Head Rotation than 2xHead (all p $<$ .001).

Cumulative Torso Rotation broken down by Direction revealed a significant Technique $\times$ Direction interaction (see \autoref{tab:SearchANOVAsByDirection}).
Post-hoc pairwise comparisons, however, showed no significant differences between individual conditions (all p = 1), indicating that Cumulative Torso Rotations were similar across the board.
The overall mean Cumulative Torso Rotation across all conditions was M = 7.116\textdegree{} (95\% CI = $\pm$1.727\textdegree{}).

\end{document}